\documentclass[12pt]{article}
\pdfoutput=1

\usepackage[pdftex]{graphicx}
\usepackage{amssymb}

\newcommand{\rf}[1]{(\ref{#1})}
\newcommand{\beq}{\begin{equation}}
\newcommand{\eeq}{\end{equation}}
\newcommand{\bea}{\begin{eqnarray}}
\newcommand{\eea}{\end{eqnarray}}

\newcommand{\e}{\mbox{e}}
\renewcommand{\d}{\mbox{d}}
\newcommand{\ra}{\rangle}
\newcommand{\la}{\langle}

\begin{document}

\begin{center}
\vspace{24pt}
{ \Large \bf A c=1 phase transition in two-dimensional 
CDT/Horava-Lifshitz gravity?  }

\vspace{24pt}

{\sl J.\ Ambj\o rn}$\,^{a,c}$,
{\sl A.\ G\"{o}rlich}$\,^{a,b},$
{\sl J.\ Jurkiewicz}$\,^{b}$
and {\sl H.\ Zhang}$\,^{b}$

\vspace{10pt}

{\small

$^a$~The Niels Bohr Institute, Copenhagen University\\
Blegdamsvej 17, DK-2100 Copenhagen \O , Denmark.

\vspace{10pt}

$^b$~Mark Kac Complex Systems Research Centre,\\
Marian Smoluchowski Institute of Physics, Jagellonian University,\\ 
Reymonta 4, PL 30-059 Krakow, Poland.

\vspace{10pt}

$^c$~Institute for Mathematics, Astrophysics and Particle Physics
(IMAPP)\\ Radbaud University Nijmegen, Heyendaalseweg 135, 6525 AJ, \\
Nijmegen, The Netherlands

}

\end{center}

\vspace{24pt}

\begin{center}
{\bf Abstract}
\end{center}

We study matter with central charge $c >1$ coupled to 
two-dimensional (2d) quantum gravity, here represented as causal dynamical 
triangulations (CDT). 2d CDT is known to provide a regularization 
of (Euclidean) 2d Ho\v{r}ava-Lifshitz quantum gravity.
The matter fields are massive Gaussian fields, where the mass
is used to monitor the central charge $c$. Decreasing the 
mass we observe a higher order phase transition between an effective 
$c=0$ theory and a theory where $c>1$. In this sense the 
situation is somewhat similar to that observed for 
``standard'' dynamical triangulations (DT) which provide a regularization
of 2d quantum Liouville gravity. However, the geometric phase observed for 
$c >1$ in CDT is very different from the corresponding phase observed for DT.

\vspace{10pt}
\vfill

{\footnotesize
\noindent
email: ambjorn@nbi.dk\\
email: goerlich@nbi.dk, atg@th.if.uj.edu.pl\\
email: jerzy.jurkiewicz@uj.edu.pl\\
email:  zhang@th.if.uj.edu.pl\\
}


\newpage

\section{Introduction}\label{intro}

Two-dimensional models of quantum gravity are useful toy models when 
it comes to study a number of conceptual problems related to a 
theory of quantum gravity: how do we define diffeomorphism invariant
observables, how do we define distance when we at the same time 
integrate over geometries etc...? Some two-dimensional models 
have the further advantage that they can be solved analytically both as 
continuum quantum field theories and as regularized ``lattice'' theories.
Quantum Liouville gravity (2d Euclidean quantum gravity) can be 
solved as a conformal field theory \cite{kpz,david,dk} and also using dynamical 
triangulations (DT) \cite{david1,kkm,adf}. 
Similarly 2d (Euclidean) quantum Ho\v{r}ava-Lifshitz 
gravity (HLG) \cite{horava} can be solved both using continuum methods 
and as a lattice theory \cite{agsw}. In both cases there seems to 
be a $c=1$ barrier: the geometries for $c<1$ and $c>1$  look 
completely different. However, it has not been easy to study 
the transition in either of the cases since no analytic solutions
exist for $c>1$ and since it is difficult to vary $c$ continuously
in numerical simulations. 

The exist many numerical studies (and a few analytic studies) of DT 
coupled to matter  in 
the $c>1$ region (for a partial list see \cite{largec}), 
and a few numerical studies of  CDT coupled to 
matter in the same region \cite{aal,agjz}. In this paper we will study the 
transition from $c < 1$ to $c>1$ in a CDT model coupled to 
four Gaussian matter fields. In order to be able to interpolate between
the two regimes we introduce a mass for the Gaussian fields. 
When the mass is large (of the order of the inverse lattice spacing)
we expect the Gaussian fields to decouple from the geometry\footnote{We work 
in the Euclidean sector of the theory. Thus no black holes
(or more precisely ``Ho\v{r}ava-Lifshitz''-like black holes) are expected
to form in this sector when we increase the mass.}
The geometry will then be that of pure 2d HLG. If the mass is zero the 
Gaussian fields will represent a conformal field theory with $c=4$.
We have already studied this system numerically \cite{agjz} and we observed 
a change of the geometry compared to the $c=0$ case.  Decreasing the mass
will  bring us from $c=0$  to $c=4$. 
On the way we will observe a phase transition between 
to the two geometric regimes. 
     
Let us briefly describe what has already been observed before the present study.
The numerical studies of 1+1 dimensional CDT are conducted using an (Euclidean)
spacetime with topology $S^1\times S^1$. The choice of a 
periodic (imaginary) time direction is mainly for numerical convenience 
and will not play a role as an indicator of finite temperature (the time 
extent can always be considered long, relative to any finite temperature 
considerations).   
In the original formulation of the CDT model in 1+1 dimensions \cite{al} 
the geometry is represented by a discretized spacetime built of triangles. 
The vertices of triangles are located at integer times, with two vertices 
at a time $t$ and one at $t\pm 1$.
Spatial slices at the discrete integer-labeled times 
then have the topology of a circle $S^1$. 
Equivalently,  one can  use the dual lattice with points in the 
centers of triangles connected by links, dual to the links of the 
triangles. The dual vertices are placed at  half-integer times. 
In the dual formulation each vertex is connected to three other vertices, 
two of which are neighbors in the same time slice and one which 
lies in the half-integer time slice positioned above or below. 
Links joining vertices with the same time index form (together with 
the corresponding vertices) ``space'' at that given time, 
and spatial topology is $S^1$. In this paper we use this dual formulation.

Without matter fields the model can be 
solved analytically \cite{al}. Let $\la n(t) \ra$ denote the average spatial 
volume measured at (half-integer) time $t$. 
If the time direction has length $L$ and the 
spacetime volume is $N$ we have $\la n(t) \ra = N/L$. The fluctuations
around this average value can also be calculated analytically. If we couple 
matter fields to the geometry, one observes the same (trivial) picture as 
long as the central charge $c\leq 1$ for the matter fields \cite{smallc}. 
However, if $c >1$ one observes a change in the behavior of the universe
\cite{aal,agjz}. If one looks at the distribution $n(t)$ 
in a single 1+1 dimensional universe generated by 
Monte Carlo simulations one observes a ``blob'' and a ``stalk''. In the 
stalk $n(t)$ is of the order of the cut-off. In the computer simulations
we do not allow $n(t)$ to shrink to zero which would result in disconnected
universes and we thus put in a lower cut off $n(t)=2$.
In the blob we have large $n(t)$'s and the average time extent 
of the blob scales as $N^{1/3}$, independent of $L$ if $L$ is larger than the
size of the blob. As a function of computer time the ``center of volume'' of
the blob is performing a random walk in the periodic time direction and
to measure average properties of the blob we have to break the translation
symmetry in our periodic discrete time. For each configuration we define 
$t=0$ as the ``center of volume'' of the blob\footnote{\label{center}
More precisely we determine the center of volume $t_{i_0}$ as follows: 
$W(i_0)$ is the minimum of the numbers  
$W(i) = \sum_{j=1}^{L} \min \{|t_i-t_j|,L-|t_i-t_j|\}\;n(t_j)$. We then 
shift the $t_i$ such that $t_{i_0}=0$, see \cite{planck} 
for a more detailed discussion in the case of higher dimensional 
CDT where the centering was first discussed.} In this way one can obtain 
the average spatial volume distribution of the blob with high accuracy:
\begin{equation}
\la n(t)\ra = \frac{2}{\pi} \alpha\, 
N^{1-1/3} \cos^2\left( \alpha \frac{t}{N^{1/3}}\right),~~~~
|t| < \frac{\pi N^{1/3}}{2 \alpha}.
\label{scaling}
\end{equation}
with $\alpha$ being a constant which depends on the central charge $c>1$  
of the matter fields, typically growing with $c$ \cite{agjz}.
The formula is only valid for the blob, i.e.\ in the $t$ range indicated
in eq.\ \rf{scaling}. For $t$ outside this range we are in the stalk and 
$n(t)$ is of the order of the cut off.   For large  spacetime volume $N$ 
the effect of the stalk can be neglected when discussing properties of 
the blob\footnote{\label{foot2}We note that the procedure of assigning $t=0$ 
to  the ``center of volume'' will introduce a bias even for 
distributions $n(t)$ where there
are no ``blobs'', as for $c<1$. In such cases we will observe an average 
distribution $\la n(t) \ra$ with a maximum at $t=0$ because of 
this bias. However, the maximum
will be very broad and there will be no stalk so the distribution is easily
distinguished from the blob-distribution \rf{scaling}. We will discuss the 
scaling of this kind of distributions in sect. \ref{large-mass}.}. 
The scaling of the 
blob as a function of the size $N$ is precisely what one expects for (a 
deformed) sphere $S^3$, $t$ being the distance from equator and we thus say
that the Hausdorff dimension  of the average two-dimension graph representing
the blob is $D_H=3$.

\section{The model}\label{model}

A massless Gaussian field has central charge $c=1$. Thus $d$ 
Gaussian fields have central charge $d$. In this paper we couple
$d$ Gaussian fields to the geometry using the CDT model.
 The scalar fields $\phi_i^{\mu},~\mu=1,\dots d$ are located at the vertices 
of the dual lattice. The  combined system of geometry and matter
is then  a statistical model described by the partition function
\begin{equation}
Z = \sum_T \frac{1}{S_T} \; \e^{-\lambda N_T}
\int \prod_{i,\mu} d\phi_i^{\mu}\e^{- S_{measure}(\phi_i^{\mu},m)}
\label{partition}
\end{equation}
where $\lambda$ is a cosmological constant, $N_T$ is the number of 
vertices in the graph dual  to the triangulation $T$ and  
$S_T$ is a symmetry factor of the graph (the order of the automorphism group 
of the graph).  The Gaussian
measure (or action) $S_{measure}$ is defined as
\begin{equation}
S_{measure}(\phi_i^{\mu}, m) = 
\frac{1}{2} \sum_{l_{ij},\mu} (\phi_i^{\mu}-\phi_j^{\mu})^2 + 
m^2 \sum_{i,\mu} \left(\phi_i^{\mu}\right)^2,
\label{action}
\end{equation}
where $l_{ij}$ is the link between vertices $i$ and $j$ and 
where we have also added a mass term to the action.

It is convenient to use $d$ massless Gaussian fields  in the simulations
if we want to study the effect of matter with central charge $d$ on the 
geometry. In contrast, using $2d$ Ising spins would require that we  first 
locate the critical point of these Ising spins coupled to the geometry and 
then conduct the simulations precisely at this critical point. Massless 
Gaussian fields are automatically critical.  Using such massless 
Gaussian fields we have measured the scaling \rf{scaling} for 
various $d >1$. 

However, it is difficult to study in detail the change of geometry
between the regime with $c<1$ and $c>1$ using massless Gaussian fields
since $d$ is an integer. In order to induce a continuous change between
the two regimes we thus introduce a mass term for the Gaussian fields. 
We start out with $d=4$ massless fields and by increasing the mass we 
will eventually for large mass have a system which effectively has $c=0$.
In principle one could obtain the same effect for multiple Ising spins
by moving gradually away from the critical point, but the procedure
is much more difficult to control numerically.

The role of the mass parameter in the action $S_{measure}$ can be
can be made clear by redefining the field variables by 
$\psi_i^{\mu} = m \phi_i^{\mu}$. Thus the action becomes 
\begin{equation}
  S_{measure}(\psi_i^\mu,m) = 
\frac{1}{2m^2} \sum_{l_{ij},\mu} (\psi_i^{\mu}-\psi_j^{\mu})^2 
+ \sum_{i,\mu} \left(\psi_i^{\mu}\right)^2
\label{action1}
\end{equation}
and the the integration measure simply redefines the cosmological constant
\begin{equation}
\prod_{i,\mu} d\phi_i^{\mu} = m^{-d N_T} \prod_{i,\mu} d\psi_i^{\mu}
\end{equation}
For large $m$ we can neglect the couplings between the 
neighboring vertices in $S_{measure}(\psi,m)$ and as a consequence,
up to a redefinition of the cosmological constant, 
we can eliminate the fields completely and obtain the pure gravity
system. For large  $m$ we thus expect 
a behavior qualitatively identical to  pure gravity  for all
geometric observables.
In the small $m$ limit we expect to observe the same kind of 
geometries as observed for massless Gaussian fields.

\begin{figure}[th!]
\begin{center}
{\scalebox{0.25}{\includegraphics{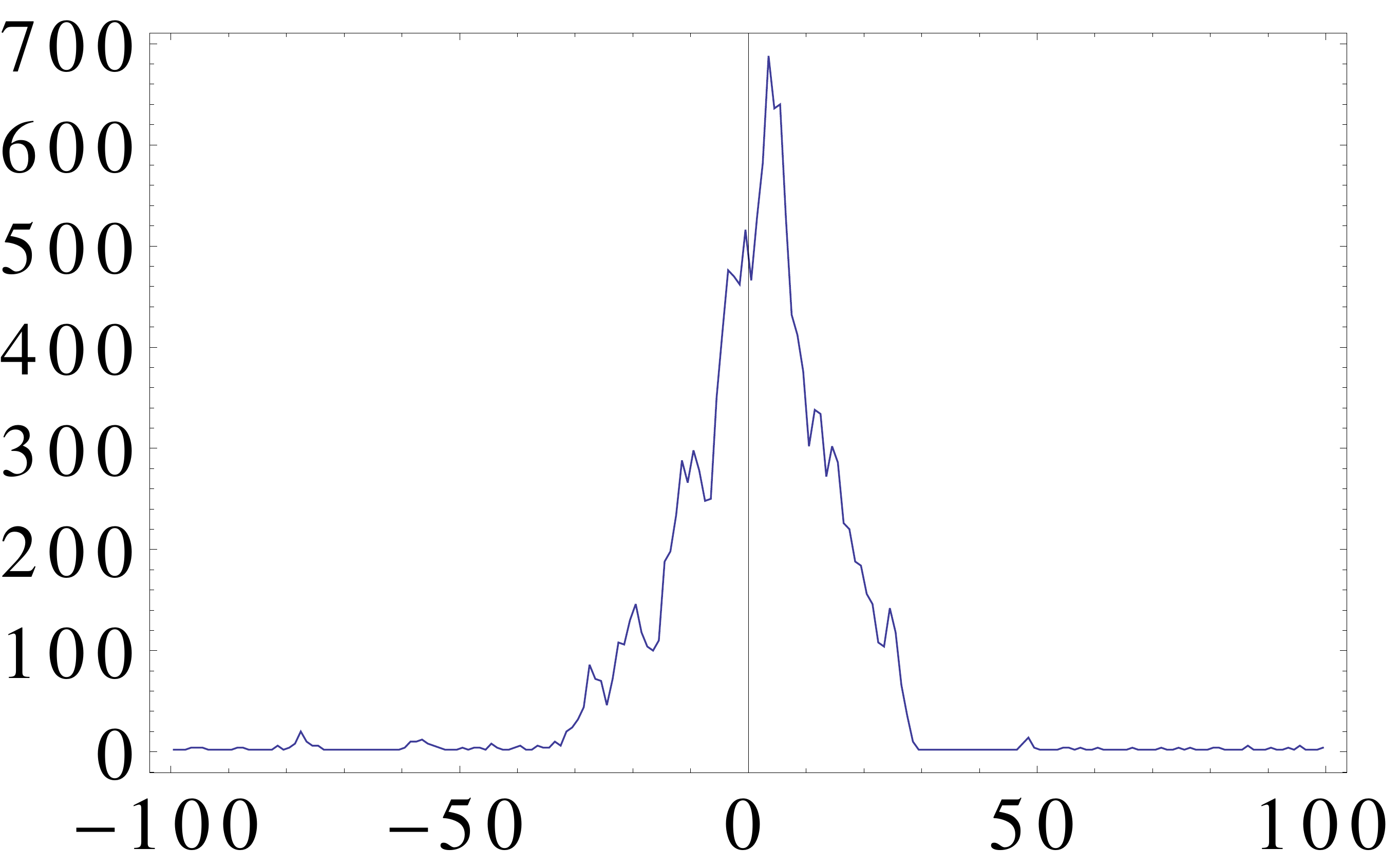}}}
{\scalebox{0.25}{\includegraphics{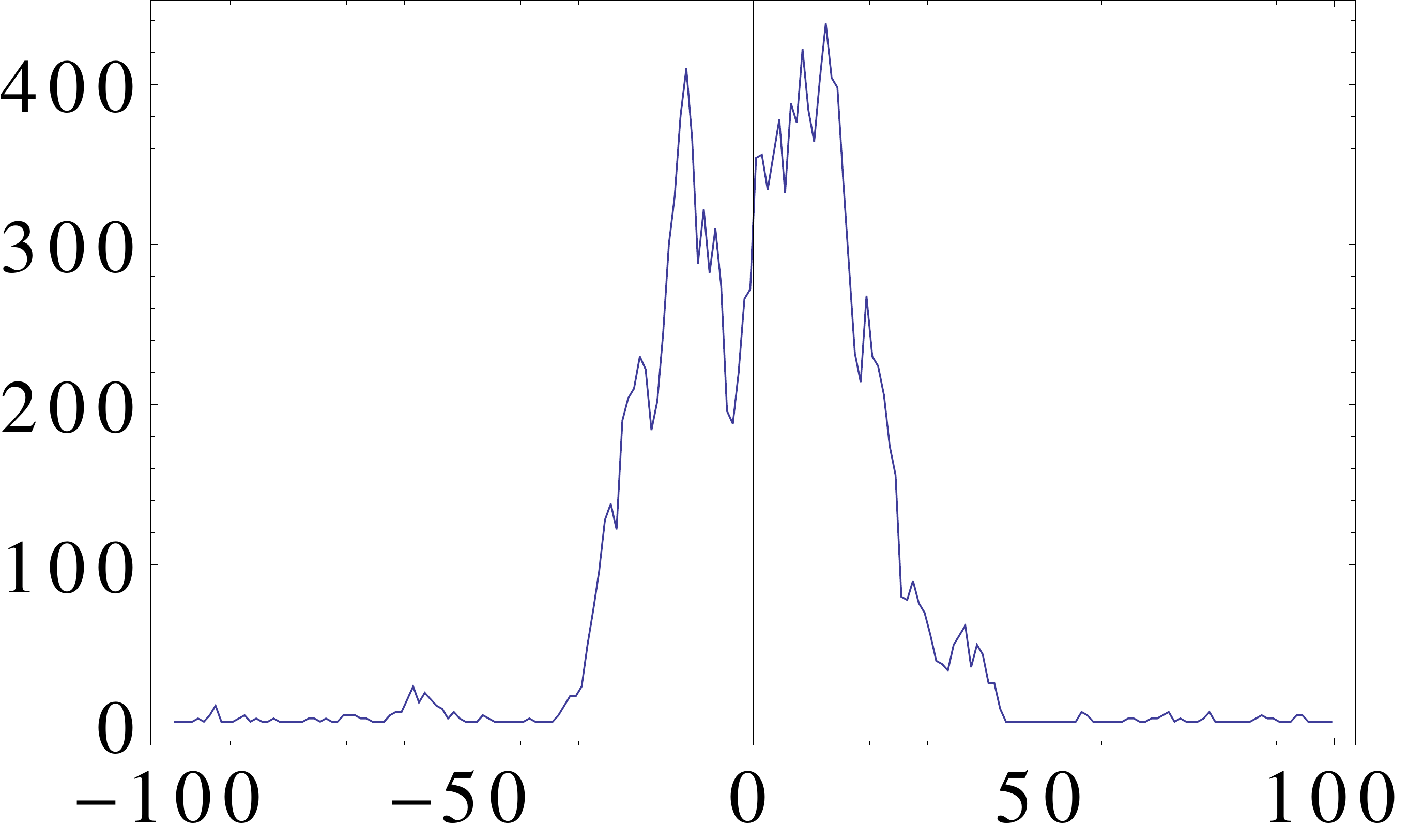}}}\\
{\scalebox{0.25}{\includegraphics{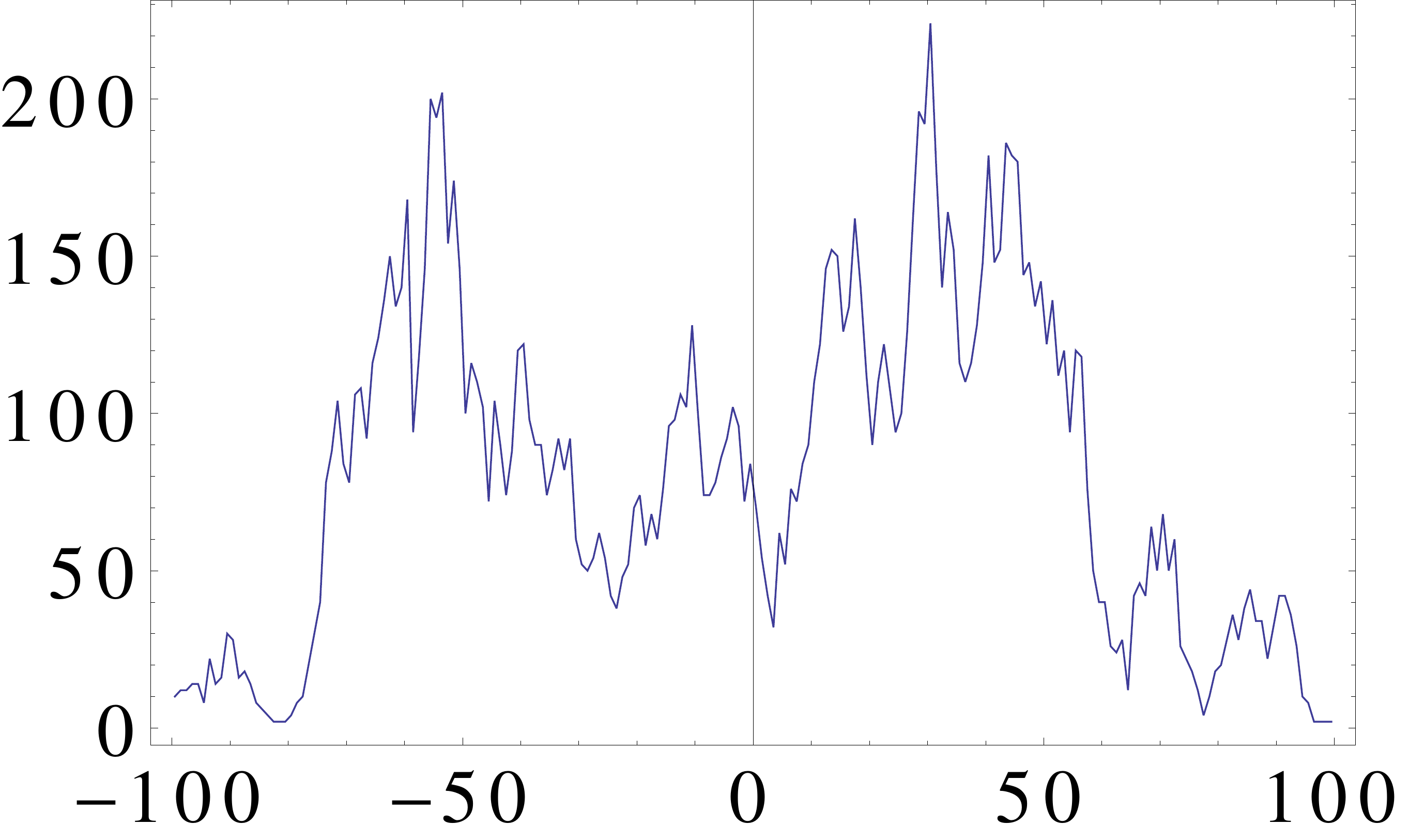}}}
{\scalebox{0.25}{\includegraphics{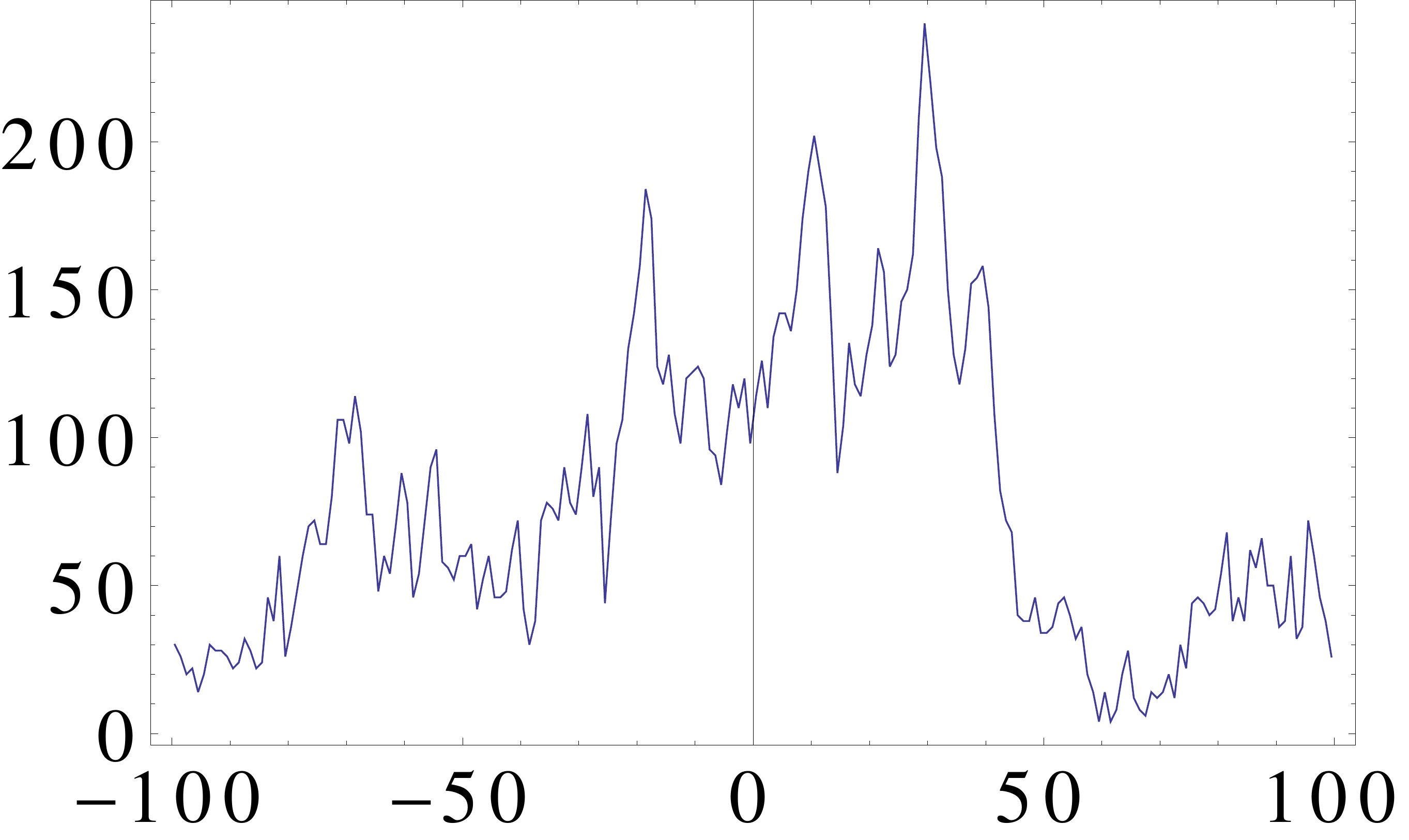}}}
\end{center}
\caption{Individual configurations $n(t,m^2)$ for $d=4$, $m^2= 0.00$ (top left),
0.05 (top right),  0.15 (bottom left) and 0.20 (bottom right). 
In all cases we center the distribution so that 
the center of volume is shifted to $t=0$.}
\label{individual}
\end{figure}

Typical configurations for various choices of masses are shown in 
Fig.\ \ref{individual} when we have four Gaussian fields. 
We see a blob for small masses,
it gets broader with increasing mass and it finally 
disappears for large masses. This will be seen 
even better when we study the average profile of the blob. 

In the next Sections we shall  quantify these effects and  try to determine a 
transition  between the two regimes as a function of the  mass parameter $m^2$.

\section{Mass dependence of the volume profiles}

\subsection{Small masses}\label{small-mass}

\begin{figure}[h]
\begin{center}
{\scalebox{0.22}{\includegraphics{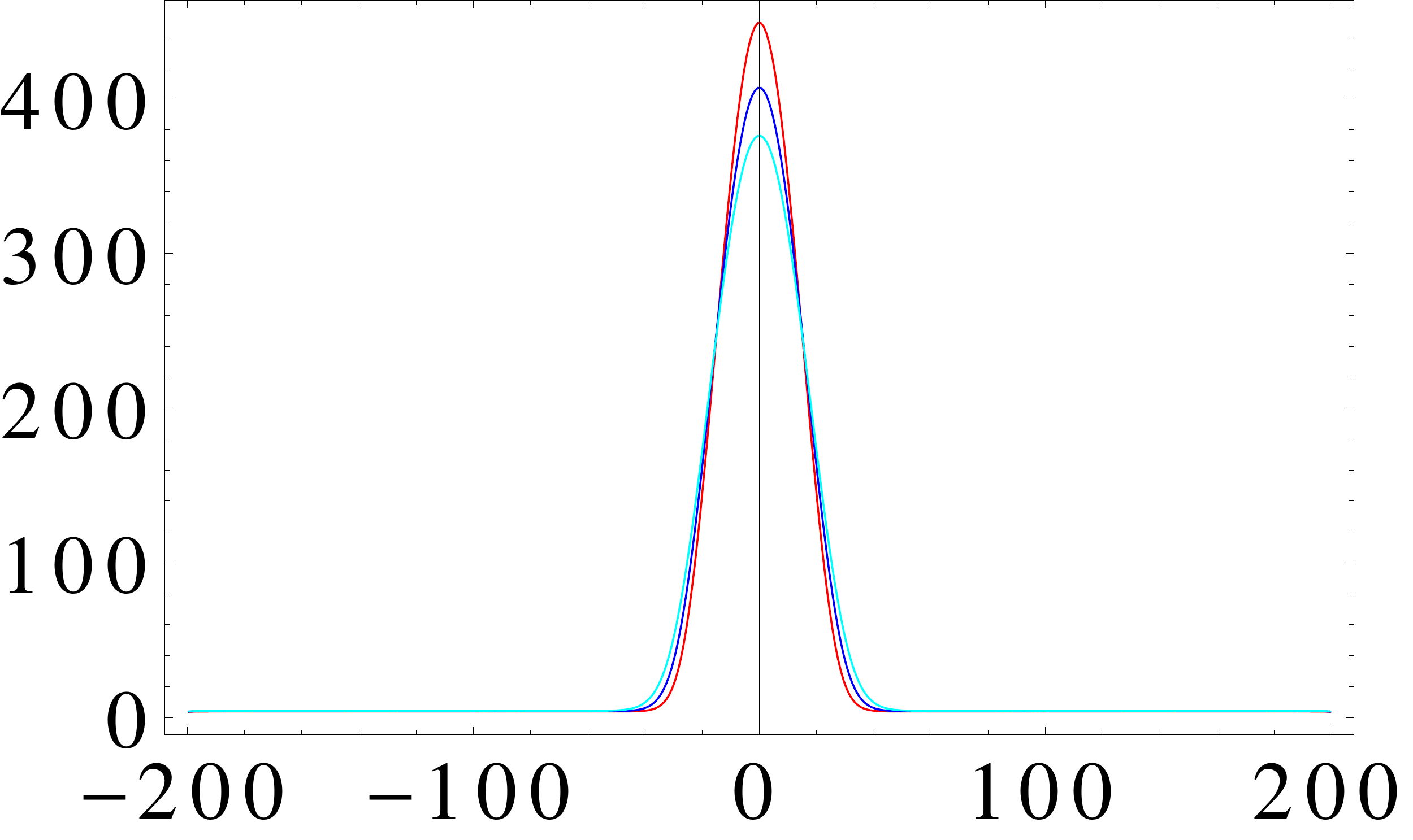}}}
{\scalebox{0.21}{\includegraphics{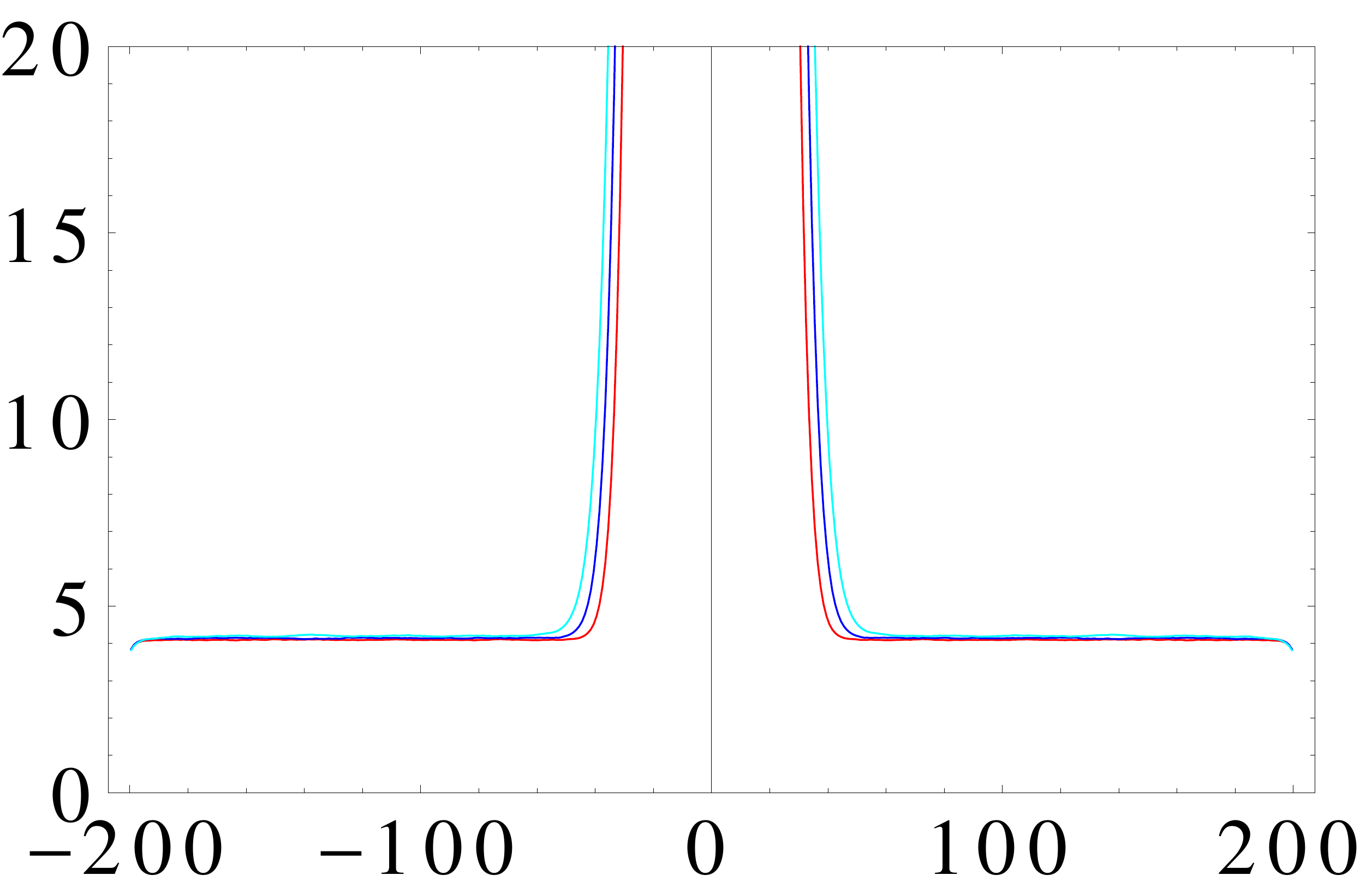}}}
{\scalebox{0.212}{\includegraphics{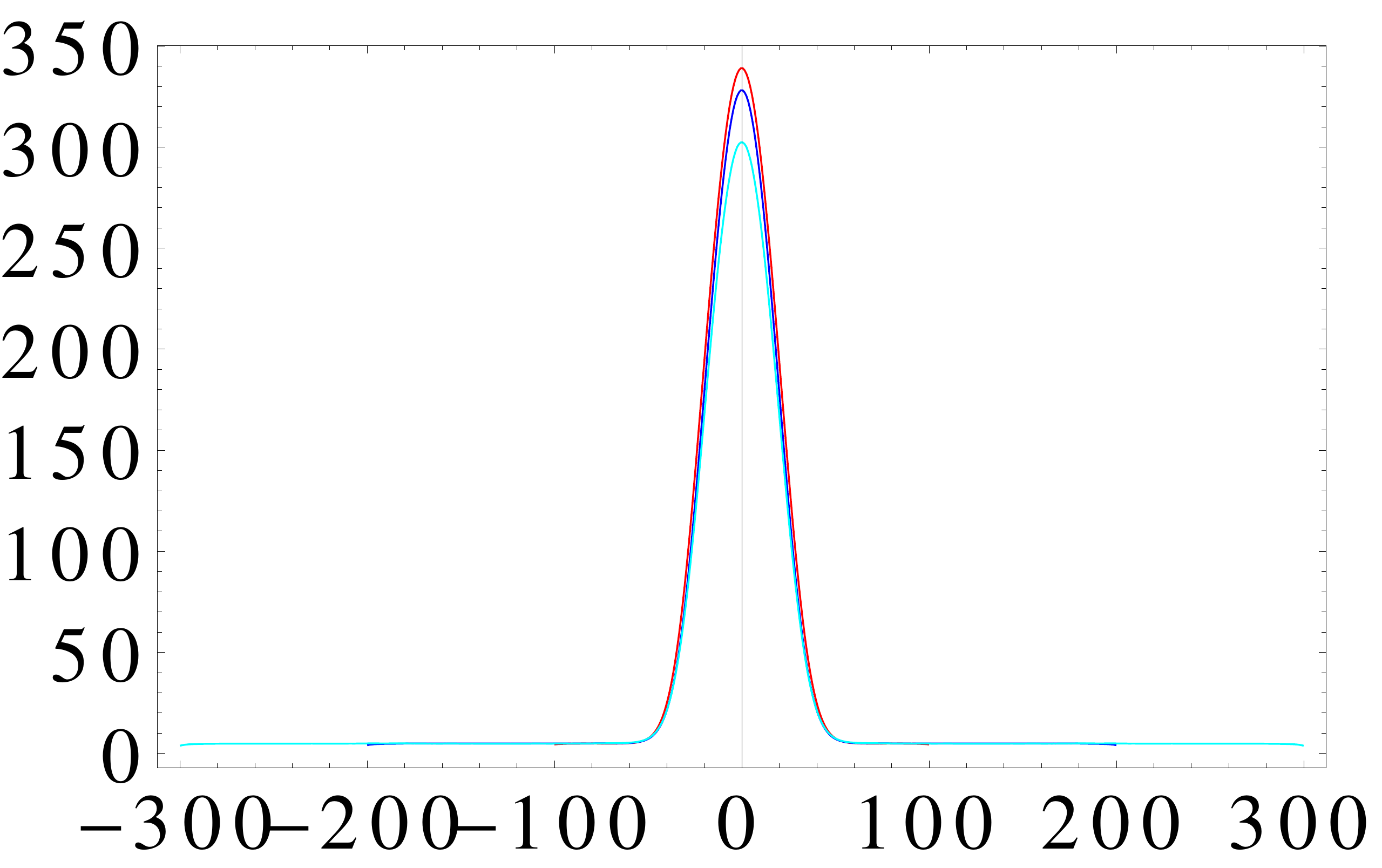}}}
{\scalebox{0.21}{\includegraphics{v-fixed-mass-001-tails.pdf}}}
\end{center}
\caption{Average spatial volume distribution $\la n(t,m^2)\ra$ for a 
fixed spacetime volume $N = 16000$ and $m^2= 0.01$ (top) and 0.05 (bottom) 
and time periods  $L=200, ~400, ~600$. On the left plots we show the 
whole range of times $-L/2 < t < L/2$. On the right plots  
we zoom in on the onset of the stalk regime in order to show that 
the stalk is  independent of $L$.}
\label{tails}
\end{figure}

For a small mass $m^2$ the average profile of spatial 
volumes $\la n(t,m^2)\ra$ contains a central blob where $\la n(t,m^2)\ra 
\gg 2$ and a stalk where $\la n(t,m^2)\ra$ is of the 
cut-off size 2. In Fig.\ \ref{tails} we show
the dependence of the volume profiles for a fixed total volume $N = 16000$ 
and a sequence of time periods $L$. We see that the thickness (the spatial
volume $\la n(t,m^2) \ra$) of the stalk does not change with $L$. 
The peaks of the blobs get slightly reduced for larger $L$ since more 
and more volume is shifted to the stalk with increasing $L$.
In addition the thickness of the  stalk does not change  
with the total volume $N$ (provided of course that $L$ is big enough to 
contain both the blob and a stalk).
$\la n(t,m^2)\ra$ in the stalk depends on the mass and grows with $m^2$. 
The values are given in the Table.\ref{height-tail} for $0 \leq m^2 \leq 0.07$.
For this range of masses they are still of the order of the cut off.

\begin{table}
\begin{center}
\begin{tabular}{|c|c|c|c|c|c|c|c|c|}
\hline
$m^2$ & 0.00 & 0.01 & 0.02 & 0.03 & 0.04 & 0.05 & 0.06 & 0.07  \\\hline
$h$ 	& 3.75 	& 3.82 	&3.96 & 4.22 	& 4.48 	& 4.84 	& 5.43	&6.29 \\ \hline	
\end{tabular}
\end{center}
\caption{Average spatial volume $\la n(t,m^2)\ra$ 
in the stalk for  $0.00 < m^2 < 0.07$}
\label{height-tail}
\end{table}

\begin{figure}[h]
\begin{center}
\includegraphics[width=0.45\textwidth]{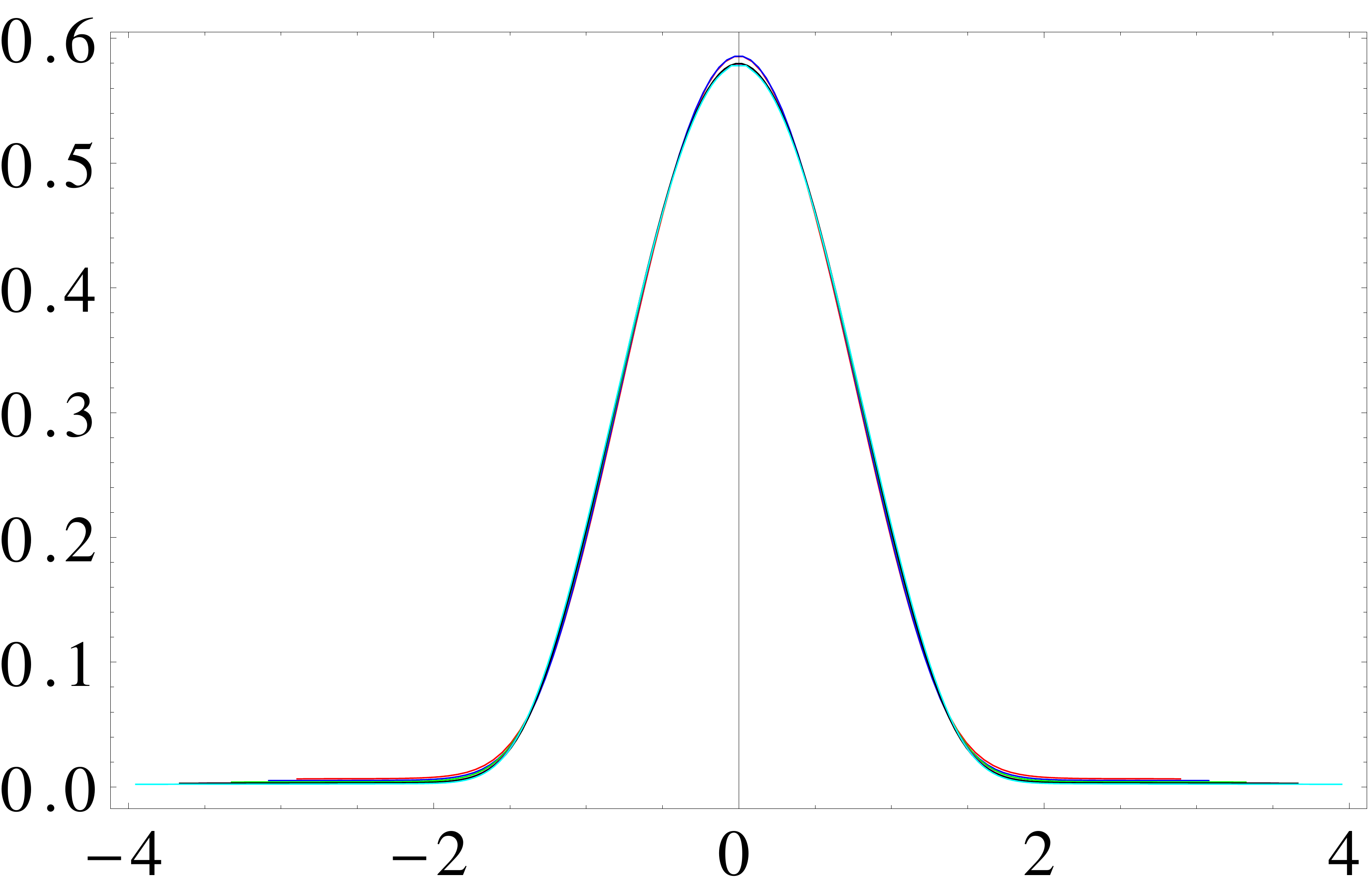}
\includegraphics[width=0.45\textwidth]{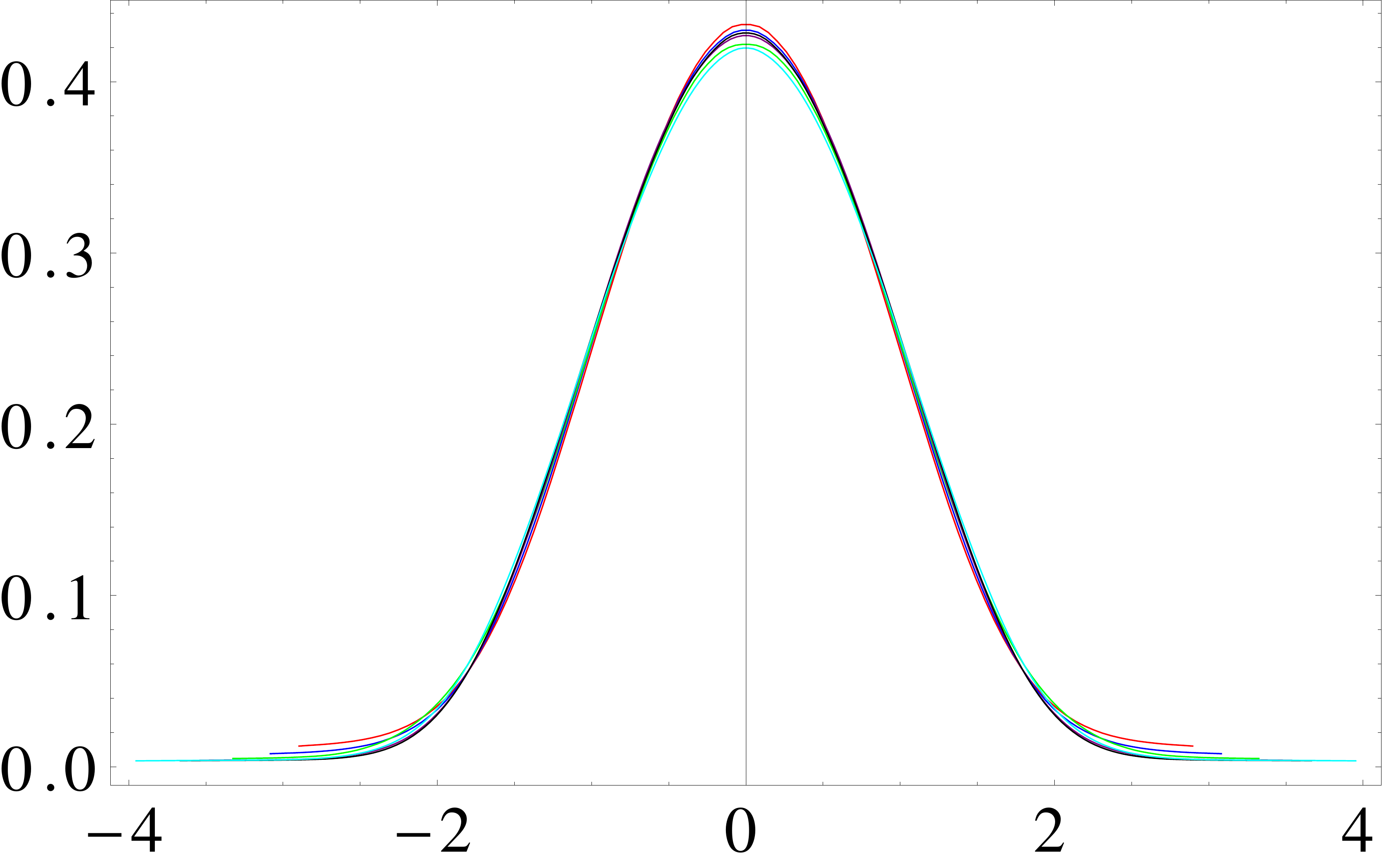}
\end{center}
\caption{The scaling function $\rho(\tau,m^2)$ which scales in the 
the blob range (but not in the stalk region), 
shown for two masses $m^2 = 0.01$ (left) 
and 0.05 (right) for a sequence of spacetime volumes $N$. Notice the change of 
scale for the different masses. The combination of volumes and 
lengths are: ($N=10000,~ L=100$), ($N=14400,~ L=120$), ($N=22500,~ L=150$),  
 ($N=32400,~ L=180$), ($N=40000, ~L=200$), ($N=62500, ~L=250$).}
\label{Haus3}
\end{figure}

For $t$ in the blob range,
$\la n(t,m^2)\ra$ scales with $N$ in a way consistent 
with a Hausdorff dimension
$D_H=3$, i.e.\ the time extent of the blob scales as $N^{1/3}$. 
In Fig.\ \ref{Haus3} 
we illustrate the scaling to  a  distribution
$\rho(\tau,m^2) = N^{1/D_H-1} n(t,m^2)$, independent of $N$, 
plotted as a function of the scaled time variable $\tau=t/N^{1/D_H}$.
The requirement of a scaling function $\rho(\tau,m^2)$ for 
different spacetime volumes  $N$ determines $D_H=3$ with good 
precision for all small values of $m^2$.  

\begin{figure}[h]
\begin{center}
\includegraphics[width=0.45\textwidth]{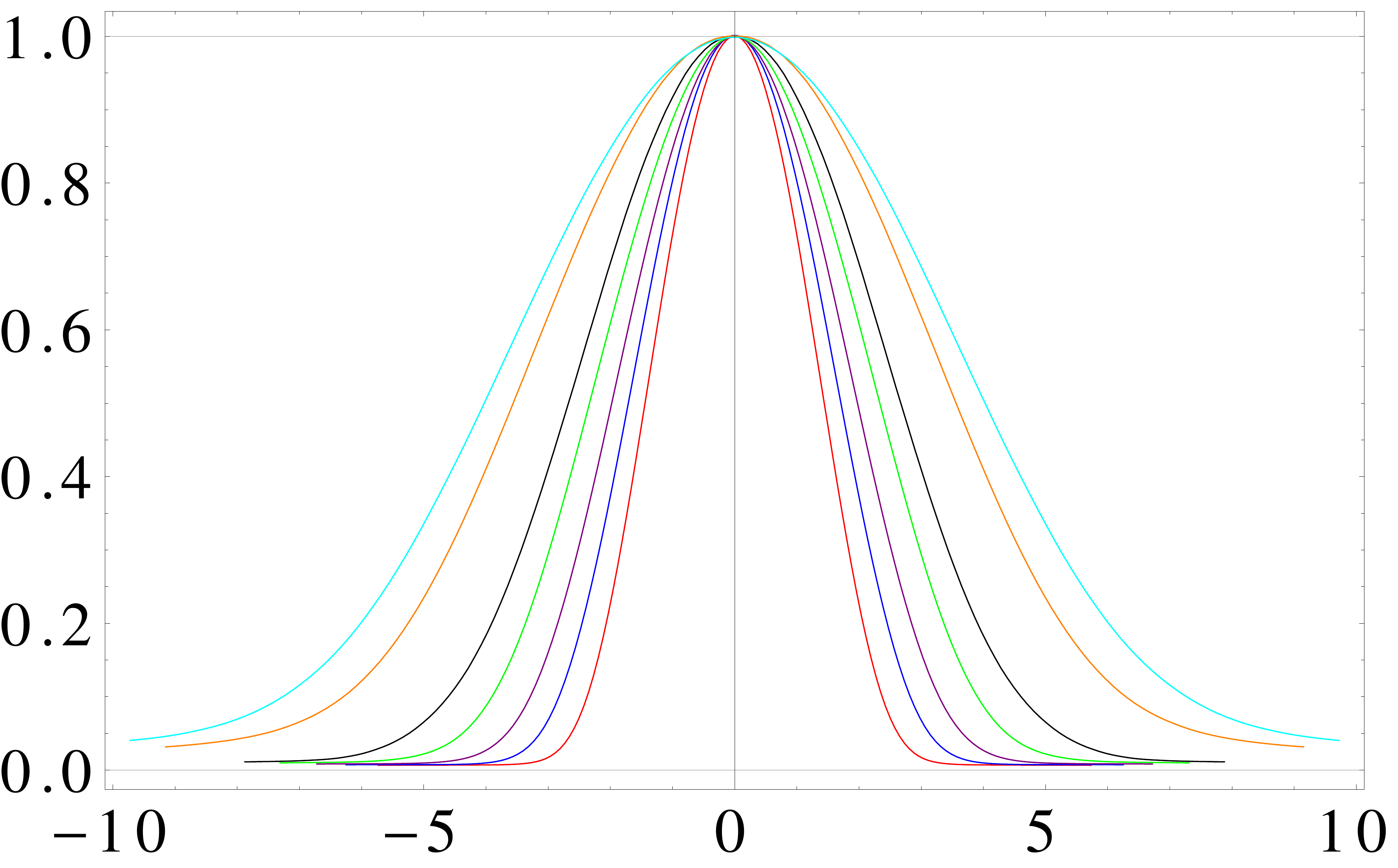}
\includegraphics[width=0.45\textwidth]{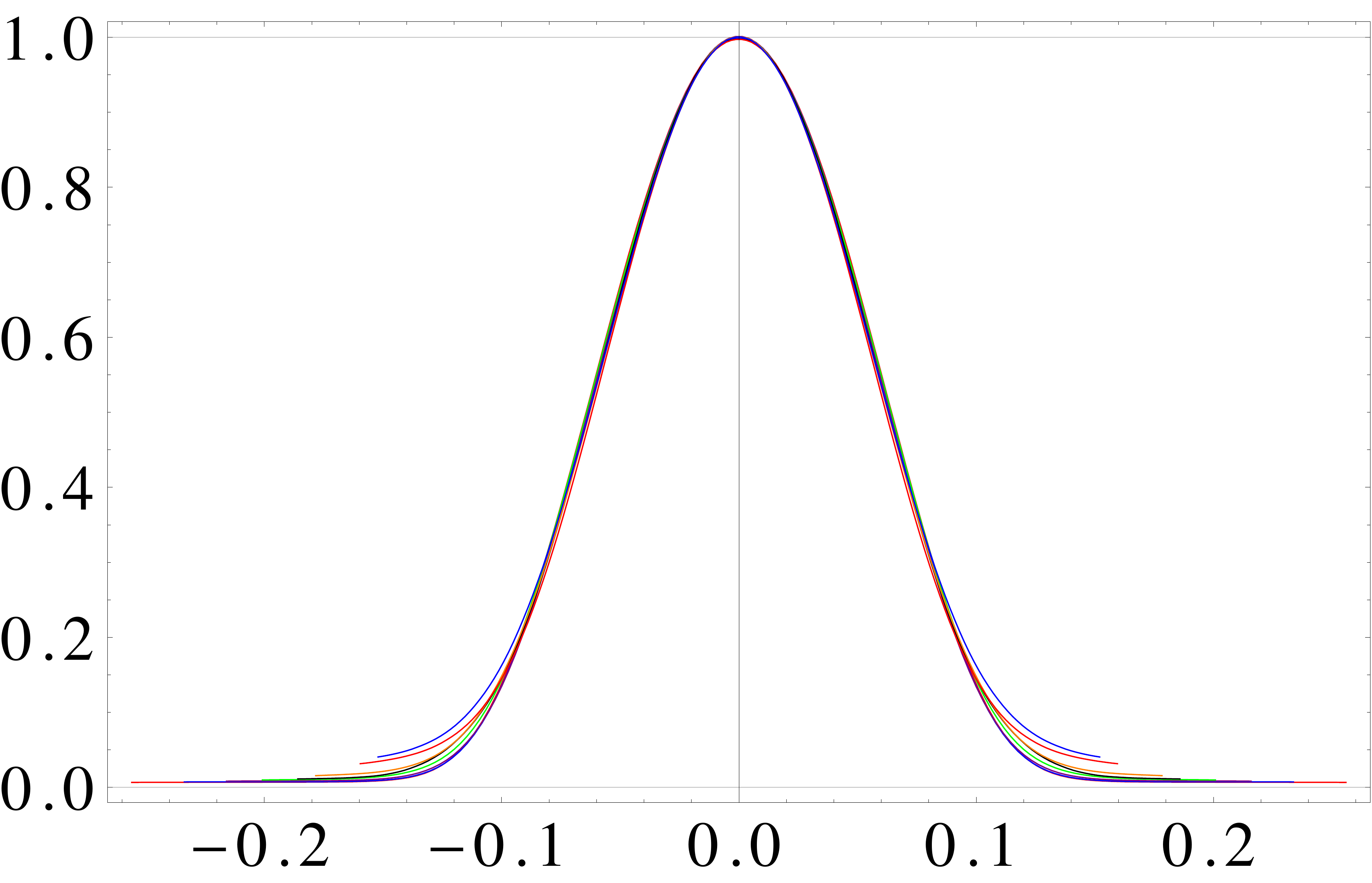}
\end{center}
\caption{Scaling for small masses $m^2\in [0,0.09]$:  
the left plot shows the distributions 
$\frac{\rho(\tau,m^2)}{\alpha(m^2)} \frac{\alpha(0)}{\rho(\tau,0)}$ 
plotted as  functions of 
$\tau = t/N^{1/3}$. The right plot shows  the same ratio plotted as 
a function of $\tilde{\tau}=\alpha(m^2) \tau$,
where the factors $\alpha(m^2)$ are determined by maximizing the 
overlap between the various curves. The  universal curve that emerges
on the right plot is $\tilde{\rho}(\tilde{\tau})$ 
up to a normalization factor $\pi/2$ 
(see eq.\ \rf{jan}).}   
\label{scaling-smallmassj}
\end{figure}

However, the universality is even larger. For small masses 
all distributions $\la n(t,m^2) \ra$ can be made to 
coincide if we, rather than scaling the time as $\tau=t/N^{1/3}$,
define a rescaled time which depends on the mass: 
$\tilde{\tau} = \alpha(m^2)\tau$  
and redefine the height of distribution accordingly
as   $\tilde{\rho}(\tilde{\tau}) = (\alpha(m^2))^{-1} N^{1/3-1}n(t,m^2)$.
A comparison of the rescaled distributions $\tilde{\rho}(\tilde{\tau})$ for 
$m^2 \in [0.01,0.09]$ is presented on  Fig. \ref{scaling-smallmassj}.
The left curves are obtained by keeping the time variable $\tau$ unchanged
but rescaling the maximum height of the curves
to the $m^2=0$ curve (which is equivalent to multiplying $\rho(\tau,m^2)$
with $\alpha(0)/\alpha(m^2)$, provided a universal 
$\tilde{\rho}(\tilde{\tau})$ exists).
The right curves are then obtained by rescaling $\tau$ to $\tilde{\tau}$
for the various curves, and in this way determining 
$\alpha(m^2)/\alpha(m^2=0)$ as the value leading to maximal overlap 
with the $m^2=0$ curve.  It thus follows from \rf{scaling} that 
\beq\label{jan}
 \tilde{\rho}(\tilde{\tau}) = \frac{2}{\pi} \cos^2{\tilde{\tau}}, 
~~~~\tau \in [-\pi/2,\pi/2].
\eeq
The $\alpha$ values drop for larger mass (see Fig.\ \ref{transiton-1j} for 
the plot of $\alpha(m^2)/\alpha(0)$), implying that the blob 
gets broader when expressed in the unscaled time-variable.
However, using the rescaled 
variable $\tilde{\tau}$  we can talk about {\it one universal 
scaling distribution \rf{jan} of spatial volumes in the ``blob'' phase}, 
independent of the  mass for $m^2 \in [0,0.09]$.

\subsection{Large masses}\label{large-mass}

\begin{figure}[h]
\begin{center}
{\scalebox{0.22}{\includegraphics{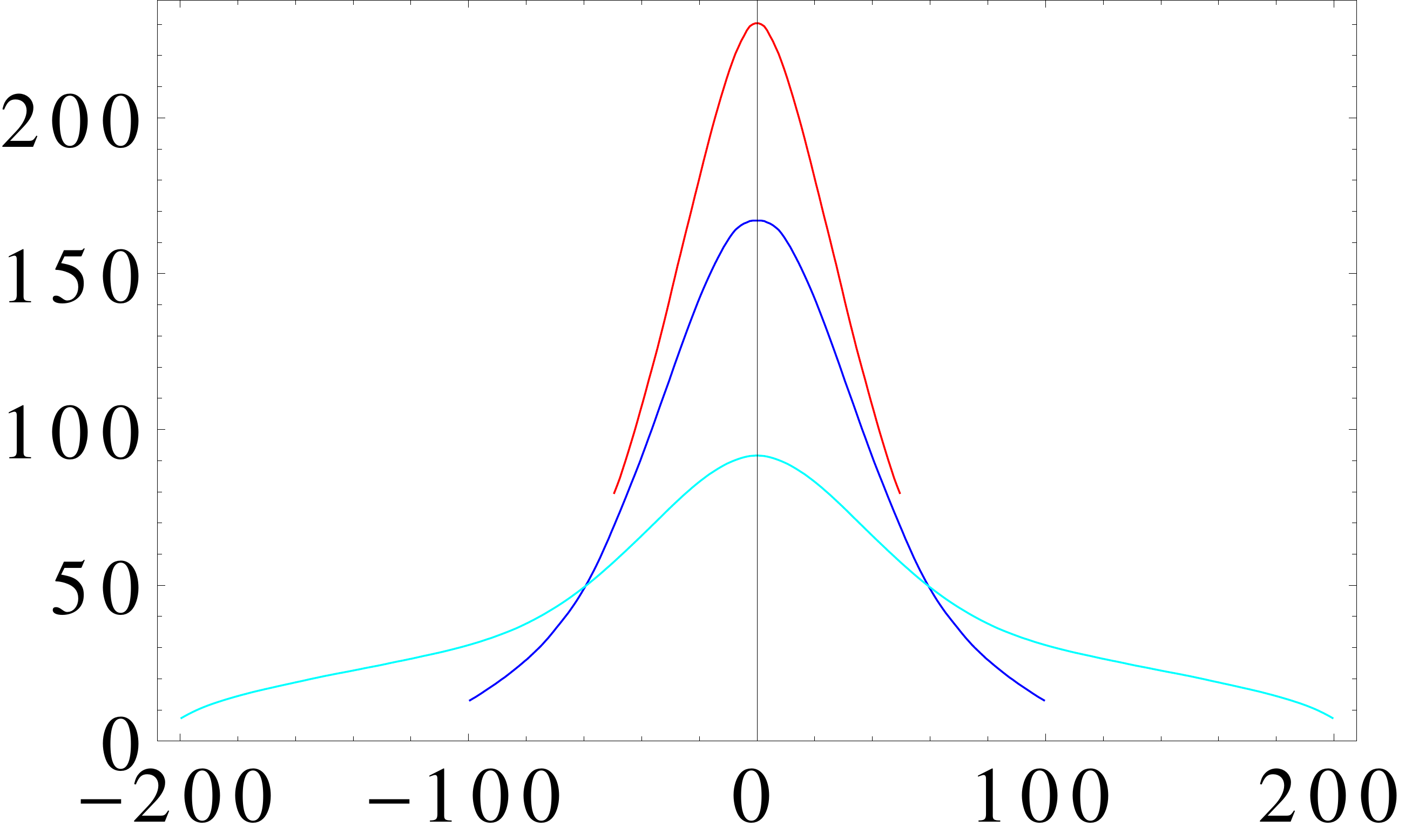}}}
{\scalebox{0.22}{\includegraphics{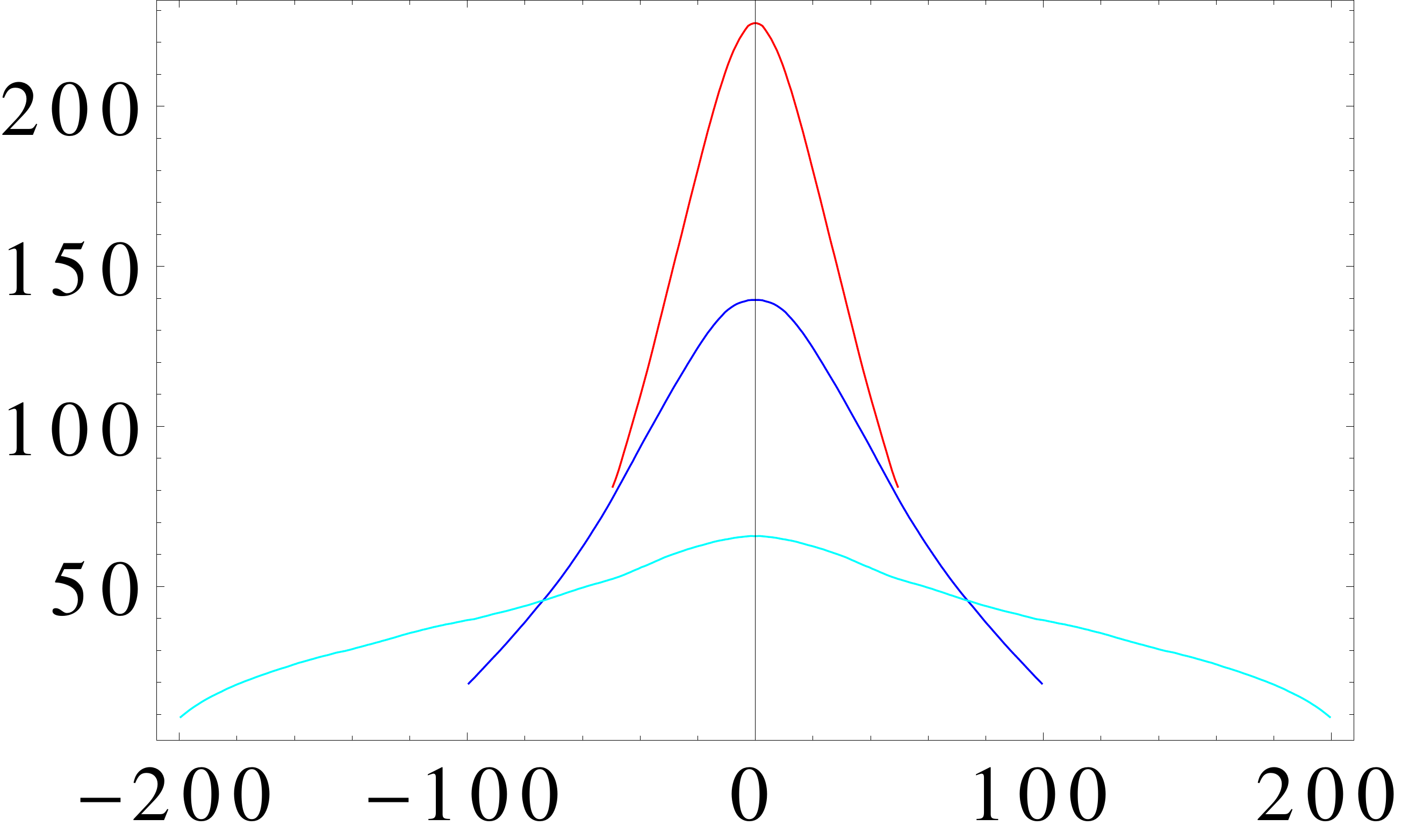}}}
\end{center}
\caption{$\la n(t,m^2)\ra$ for $m^2=0.15$ (left) and 0.20 (right) for different 
$L$ and a fixed  spacetime volume $N = 16000$.}
\label{big-massj}
\end{figure}

The behavior is different for masses $m^2 \geq 0.15$.
As was explained above we
expect for large masses that $\la n(t,m^2)\ra$  will be  qualitatively 
similar to the pure gravity case, where it is known analytically  
that any scaling should correspond to a 
Hausdorff dimension $D_H = 2$. 
In our approach we use the same method 
to center the volume of individual configurations as we used 
in the case where we observed a genuine blob (see footnote \ref{center}).
As a consequence we see an artificial maximum around time $t=0$, as 
already mentioned in footnote \ref{foot2}. 
The stalk is absent, and the distributions have  triangular
shapes, with the height depending on the assumed period $L$ as $1/L$. 
Thus 
\beq\label{jan2}
\la n(t,m^2) \ra = \frac{N}{L} f(t/L,m^2),~~~
\int_{-1/2}^{1/2} f(x,m^2) dx \approx 1.
\eeq 
This is illustrated in  Fig.\ \ref{big-massj}.

\begin{figure}[h]
\begin{center}
\includegraphics[width=0.45\textwidth]{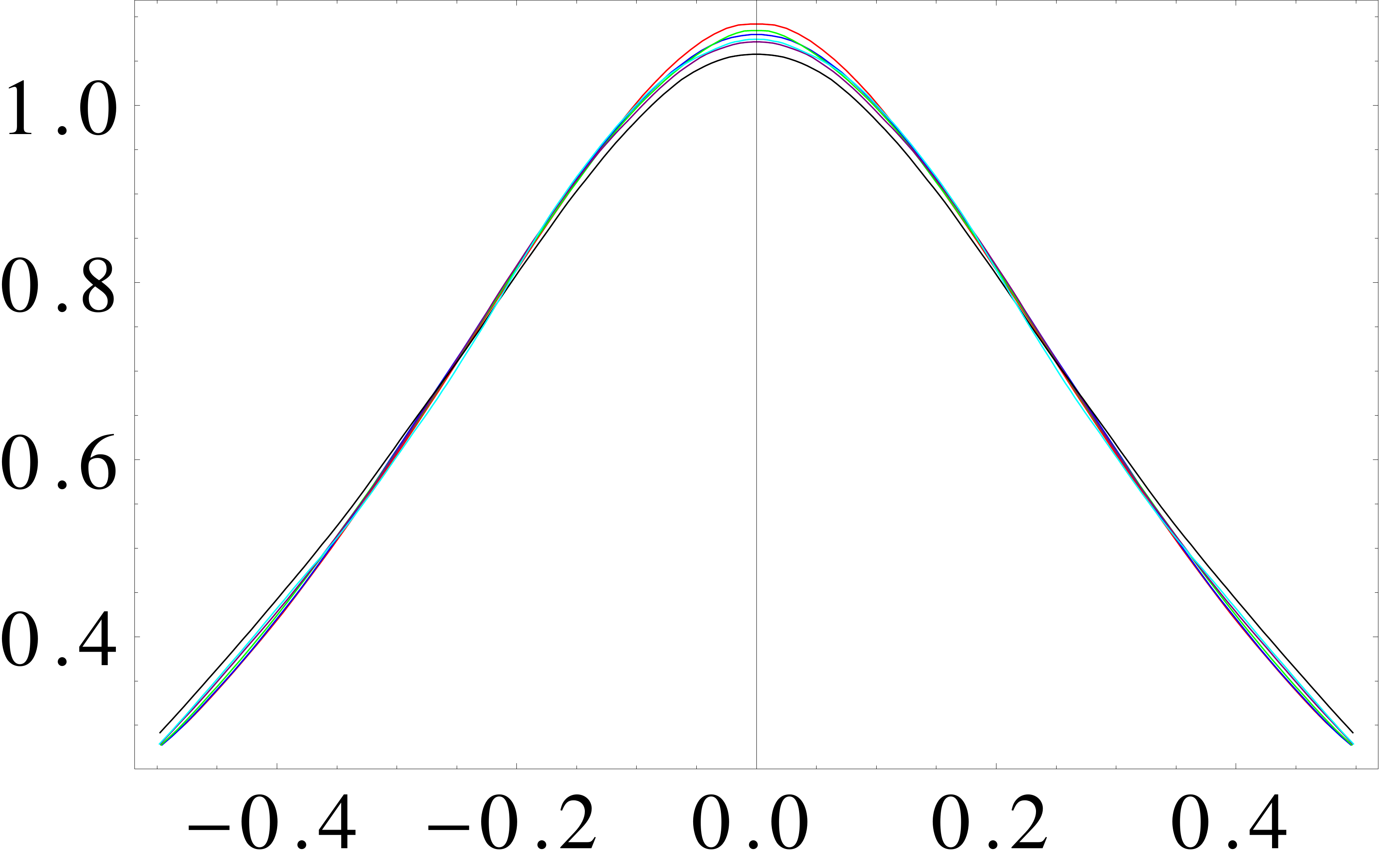}
\includegraphics[width=0.45\textwidth]{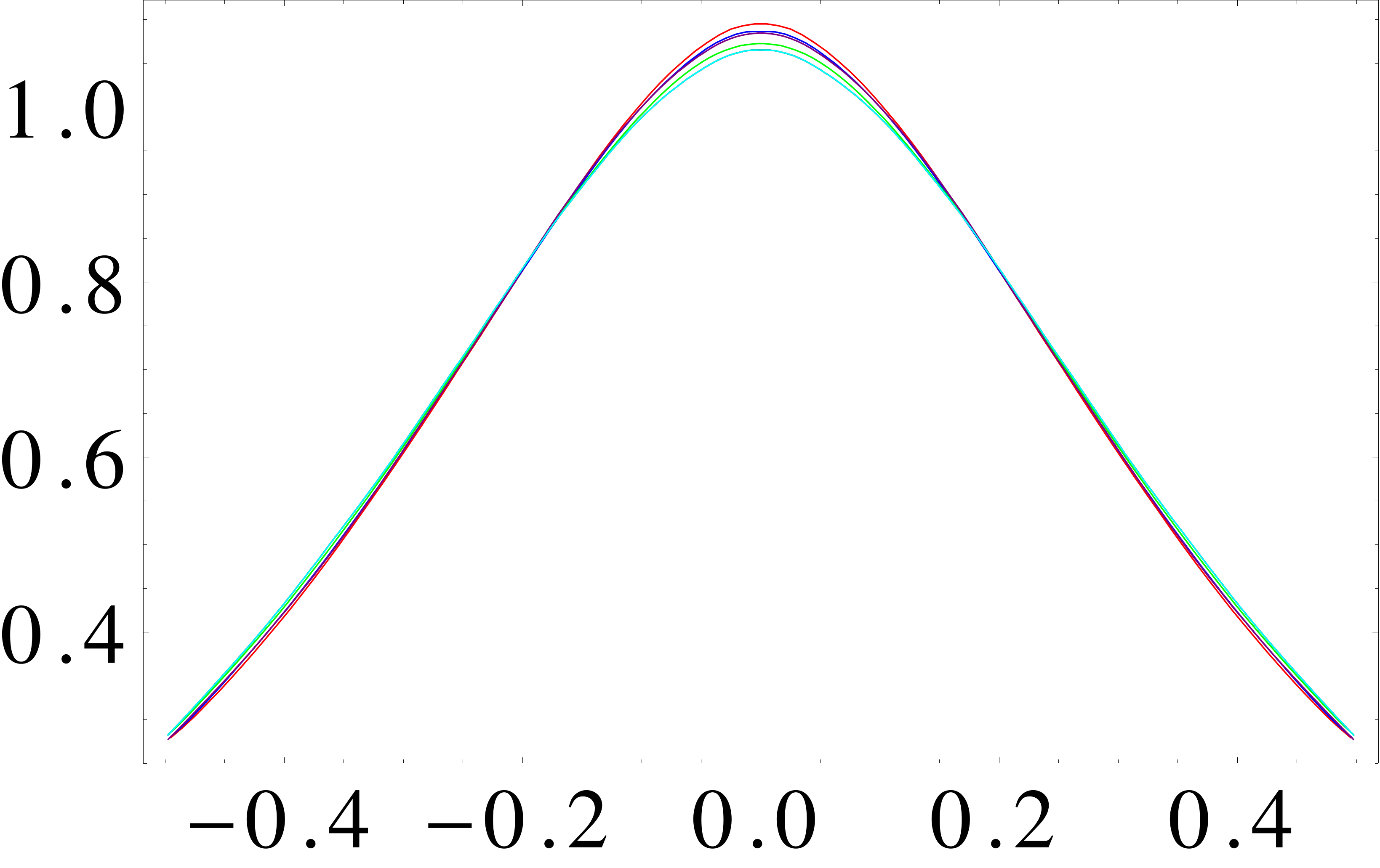}
\end{center}
\caption{The scaling function $f(\tau,m^2)$, $\tau = t/L$ as it appears
by collapsing the various graphs $N^{1/D_H-1}\la n(t,m^2)\ra$ 
for $m^2=0.15$ (left) 
and 0.20 (right), with $D_H=2$. 
The combination of volumes and lengths are: 
($N=10000,~ L=100$), ($N=14400,~ L=120$), ($N=22500,~ L=150$), 
($N=32400, ~L=180$), ($N=40000,~ L=200$),($N=62500,~ L=250$).}
\label{scaling-bigj}
\end{figure}

If we want to look for  scaling behavior of $\la n(t,m^2)\ra$ 
when changing  $N$, we have to change the length of the 
time period $L$ simultaneously as $L \propto N^{1/2}$ 
since there is no stalk.  For the choice $L = \sqrt{N}$ eq.\ \rf{jan2} reads
\beq\label{janx}
\la n(t,m^2) \ra = N^{1/2} f(t/N^{1/2},m^2).
\eeq 
For each value of  $m^2 \geq 0.15$ we can extract a scaling
function $f(\tau,m^2)$, $\tau = t/N^{1/2}$, by varying $N$,  
as illustrated in Fig.\ \ref{scaling-bigj}.
When comparing \rf{janx} with the general scaling form 
$N^{1-1/D_H} f(t/N^{1/D_H},m^2)$ we see that the observed scaling indeed is 
compatible with $D_H =2$.

\begin{figure}[h]
\begin{center}
\includegraphics[width=0.45\textwidth]{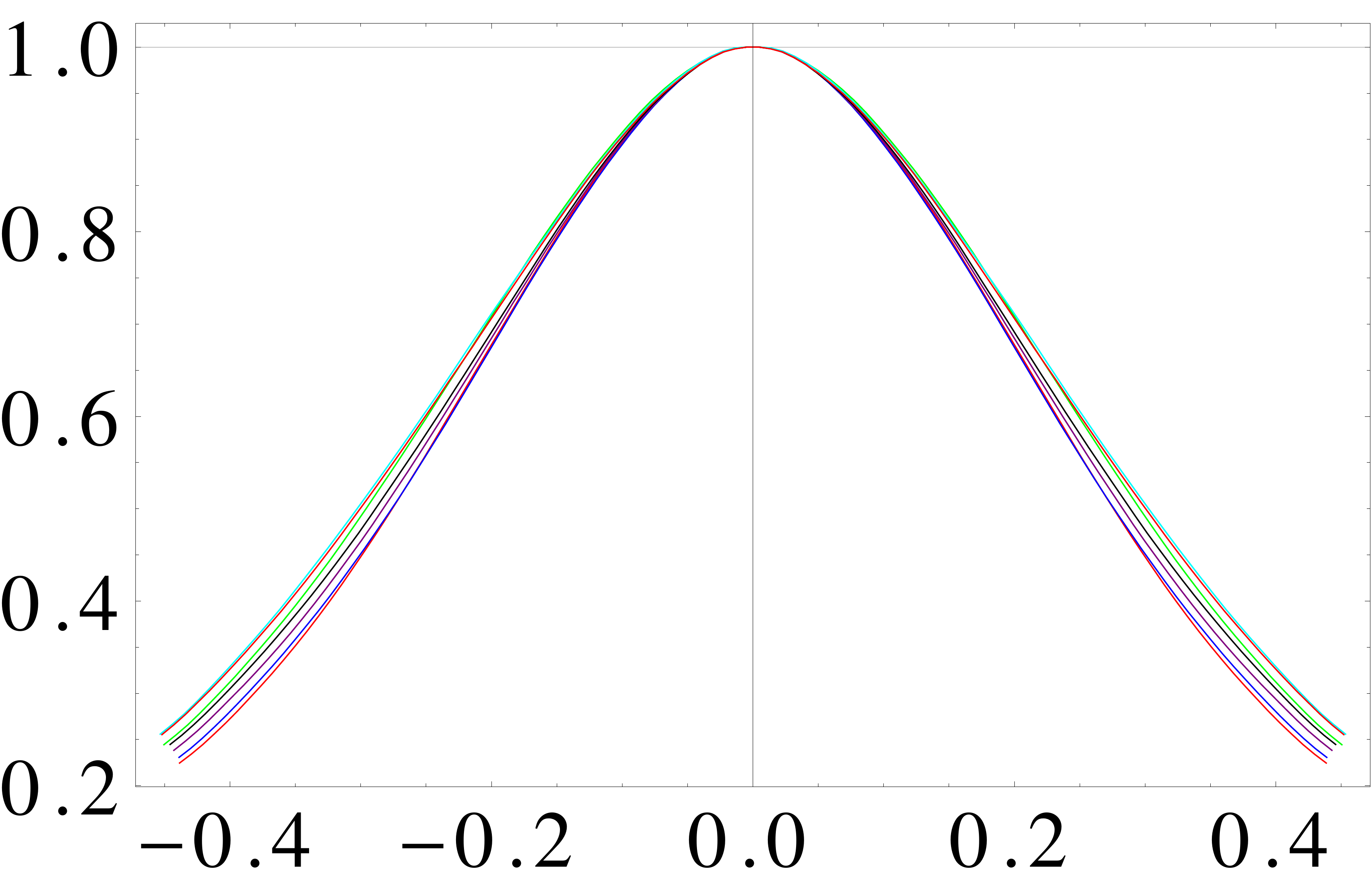}
\includegraphics[width=0.45\textwidth]{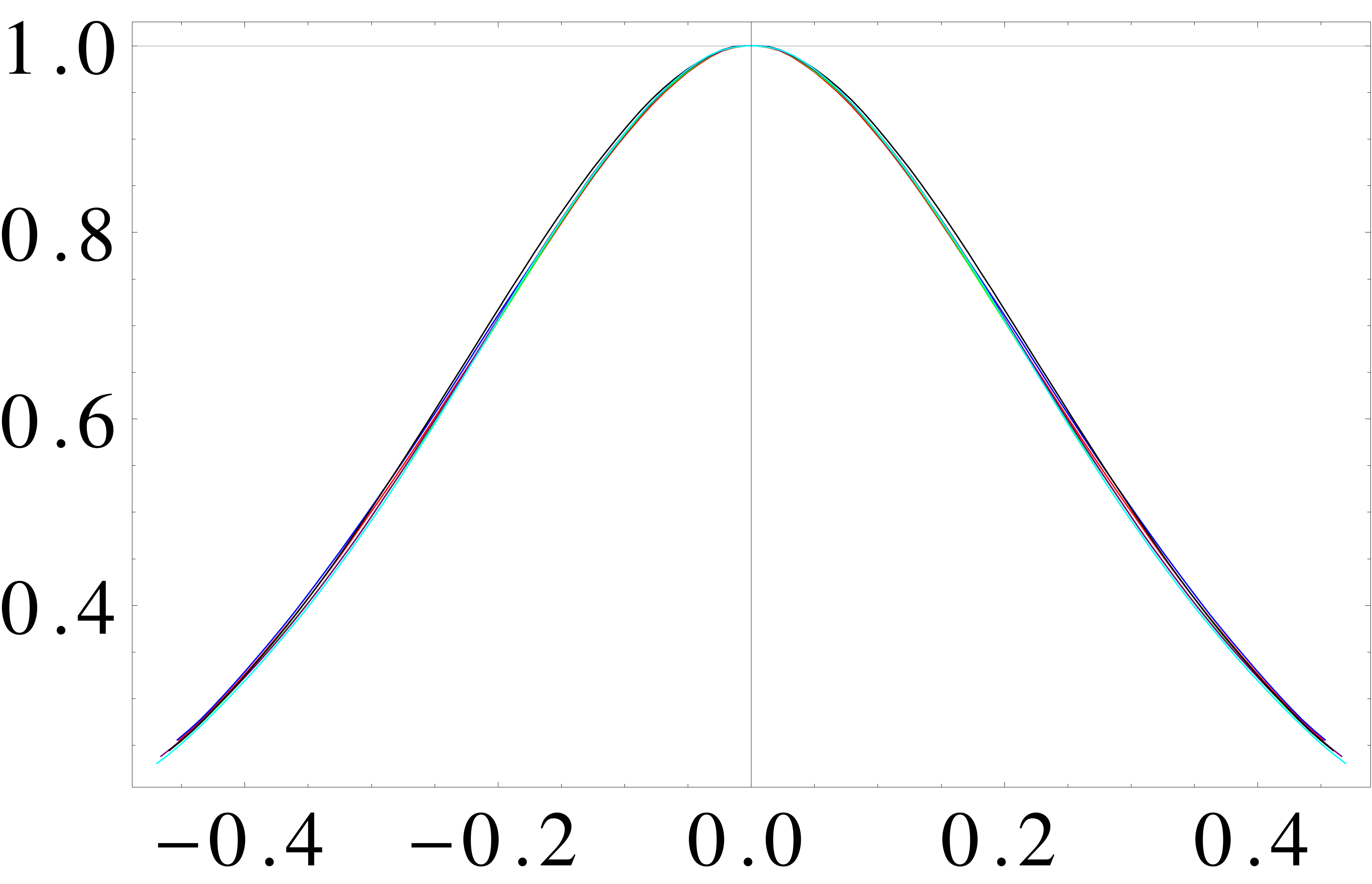}
\end{center}
\caption{The function $\hat{f}(\tau,m^2)$ (left plot) and the universal 
scaling function $\tilde{f}(\tilde{\tau})$ (right plot) for 
$m^2\in [0.12,0.18]$.}
\label{scaling-bigmassj}
\end{figure}

\begin{figure}[h]
\begin{center}
\includegraphics[width=0.75\textwidth]{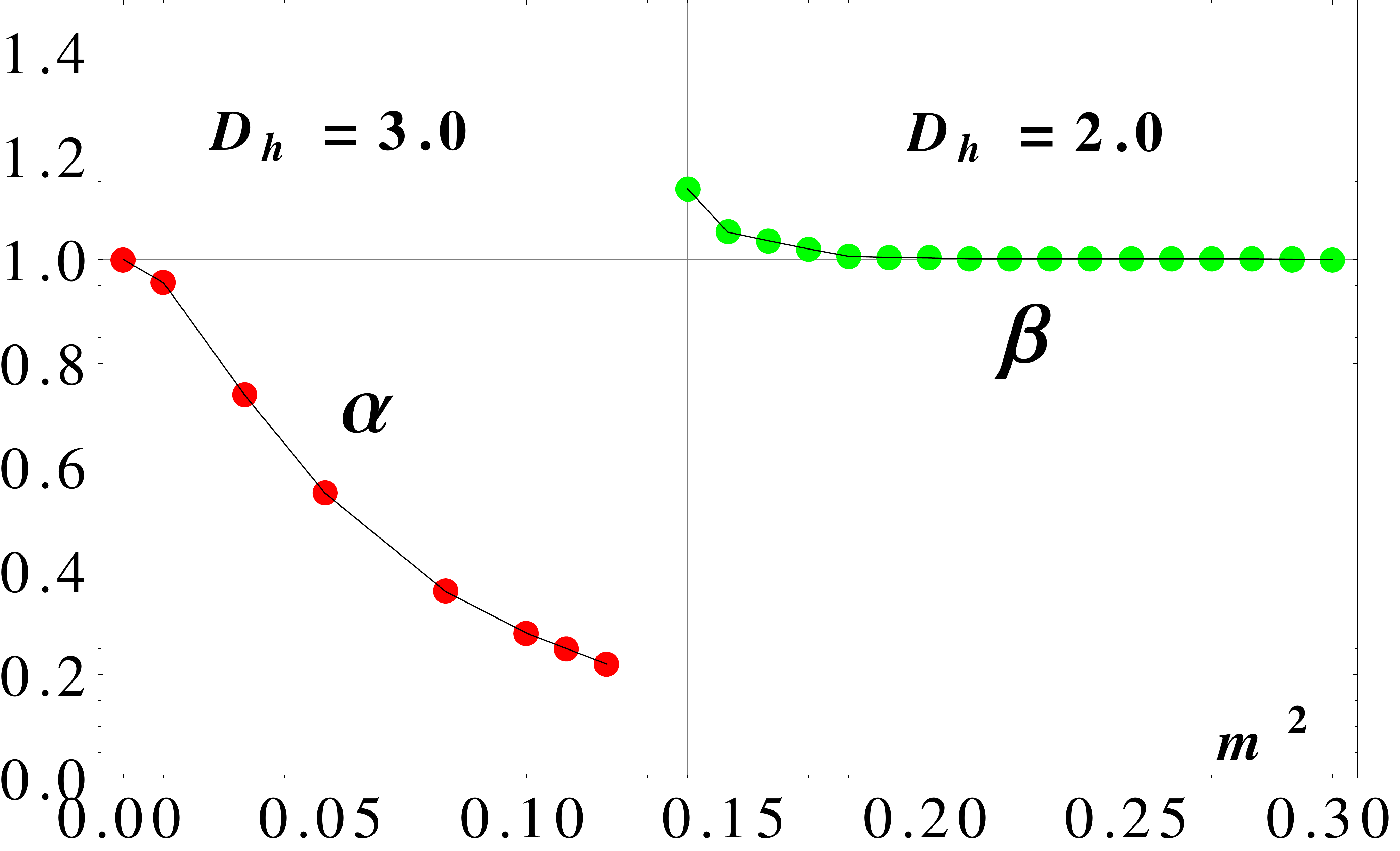}
\end{center}
\caption{Possible range of the phase transition: 
$\alpha(m^2)/\alpha(0)$ is the scale factor for the small masses  
and  $\beta(m^2)/\beta(\infty)$ the scaling factor for  large masses. 
The Hausdorff dimension is respectively  $D_H =3$ and $D_H =2.0$ .}
\label{transiton-1j}
\end{figure}

In the same way as we did for the small masses, we now try to find a universal 
scaling  of $\la n(t,m^2)\ra $ for all large masses. We 
construct the universal function in two steps, starting 
from the scaling functions $f(\tau,m^2)$, $\tau=t/N^{1/2}$, we already
have available for each $m^2$. First we scale these functions such that 
they agree at $\tau=0$ using $f(\tau,m_{max}^2)$ as reference, i.e.
\beq\label{jan4}
\hat{f}(\tau,m^2) = \frac{f(0,m^2_{max})}{f(0,m^2)} \, f(\tau,m^2),
\eeq
 The result is shown in the left plot in Fig.\ \ref{scaling-bigmassj}. 
Then we try to rescale the time variable as we did for the small masses:
$\tilde{\tau} = \beta(m^2)\,t/L$, where the function $\beta(m^2)$ is 
determined to ensure maximal overlap. This results in our universal 
scaling function
\beq\label{jan5}
\tilde{f}(\tilde{\tau}) = \hat{f}(\tau,m^2),~~~~\tilde{\tau} = \beta(m^2)\tau
\eeq
The result is shown in the right
plot in Fig.\ \ref{scaling-bigmassj} and the function $\beta(m^2)$ is shown
Fig.\ \ref{transiton-1j}. Note that this procedure results in 
overlapping graphs, but not in the same range of the common scaling 
variable $\tilde{\tau}$. This is in contrast to the situation for 
small masses, the reason being that for small masses the original $t$
range was a ``physical'' range, namely the time extent of the blob, which
was a function of $m^2$ and our rescaling $\tilde{\tau} = \alpha(m^2)\tau$
made the time extent of the blobs agree and thus, if we had first adjusted
the maximum height of the blobs, would also ensure the collapse of 
the blobs to a universal curve (assuming such one exists). In the large mass 
case we have no blob and by definition the range of $t$ is from $-L/2$ to 
$L/2$. Thus, by in addition choosing $L=N^{1/2}$ and $\tau = t/N^{1/2}$ we 
have also chosen the range of $\tau$ to be identical for the various 
large masses. Redefining $\tau$ to $\tilde{\tau}$ will then change the 
range of $\tilde{\tau}$ for the various masses and we can only talk 
about the overlap of functions in the common $\tilde{\tau}$
region. One could have compensated for this by choosing from the outset
different $L$'s according to $L = L/\beta(m^2)$ and one would have 
had a starting point similar to the small mass case where the range of $t$ was 
mass dependent. From this point of view it makes sense to talk about 
{\it one universal scaling function associated with $\la n(t,m^2)\ra$ also
in the large mass regime}, and this scaling function can be extracted from 
pure CDT without matter fields, which is the limit of $m^2 \to \infty$.
From Fig.\ \ref{transiton-1j} it is seen that  the function $\beta(m^2)$
is 1 for $m^2 > 0.18$. $m^2= \infty$ thus effectively starts at $m^2=0.18$.
On the fig.\ref{transiton-1j} we show values of 
$\alpha(m^2)/\alpha(0)$ and $\beta(m^2)/\beta(\infty)$ 
as functions of $m^2$. For 
$m^2 \in [ 0.10,0.14]$ there is a cross over between the  
two well defined regimes corresponding to $D_H = 3$ and $D_H=2$, respectively. 
In this range none of the fitting prescriptions described above work and 
using scaling arguments alone do not allow us to determine if there 
is a genuine phase transition or just a rapid cross over between the 
$D_H=3$ and the $D_H=2$ regions of $m^2$.

\section{Study of the phase transition}\label{position}

In order to study better the change from $D_H=3$ to $D_H=2$ we 
introduce the so-called volume-volume correlator 
$\langle \mathrm{corr}(\Delta)\rangle $, where 
$\mathrm{corr}(\Delta)$ is defined for individual configurations as
\begin{equation}
\mathrm{corr}(\Delta)=\sum_{i=1}^L n(t_i)n(t_i+\Delta)
\label{corrr}
\end{equation}
A great advantage of using the correlation function \rf{corrr} is that 
one does not need to identify and to center the blob and it
is well defined even if there is no blob.

\begin{figure}[h]
\begin{center}
{\scalebox{0.55}{\includegraphics{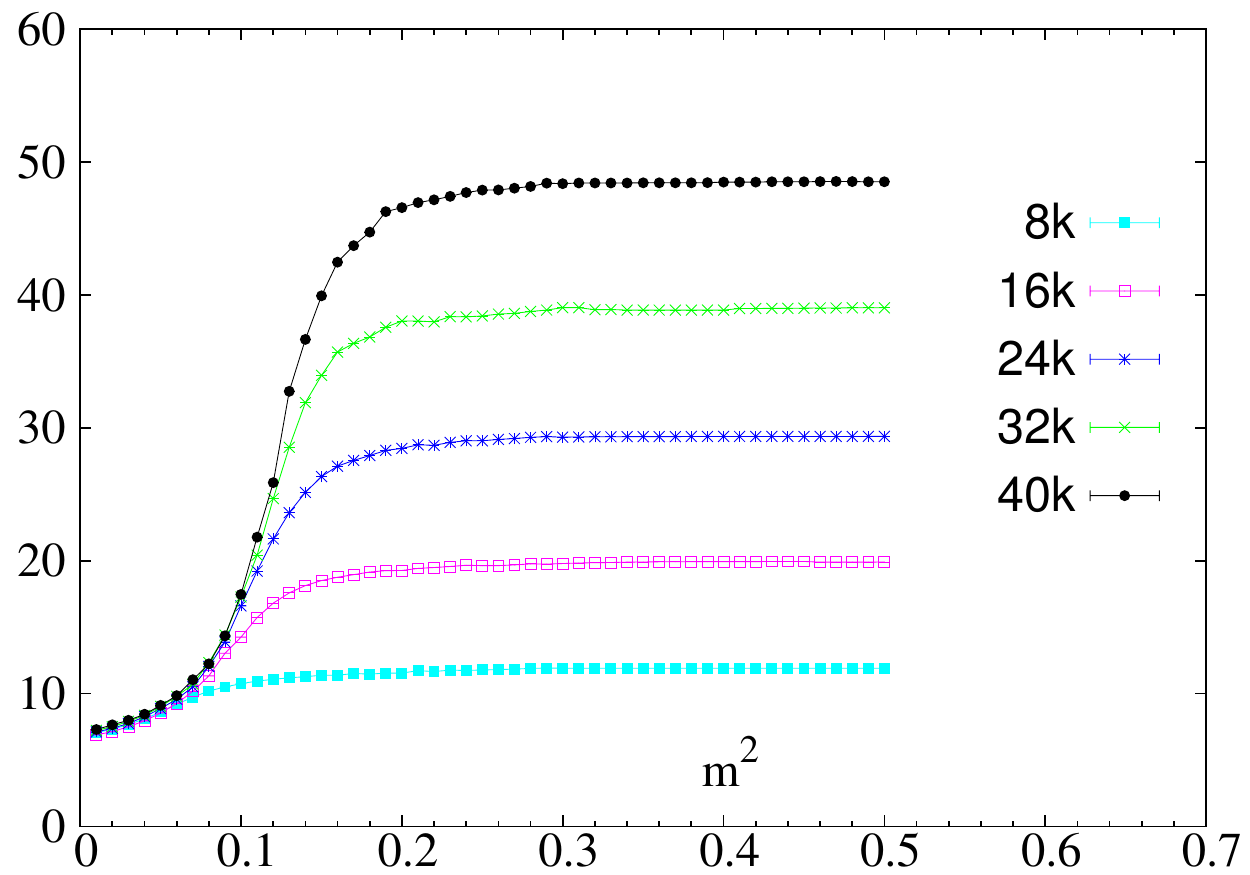}}}
{\scalebox{0.55}{\includegraphics{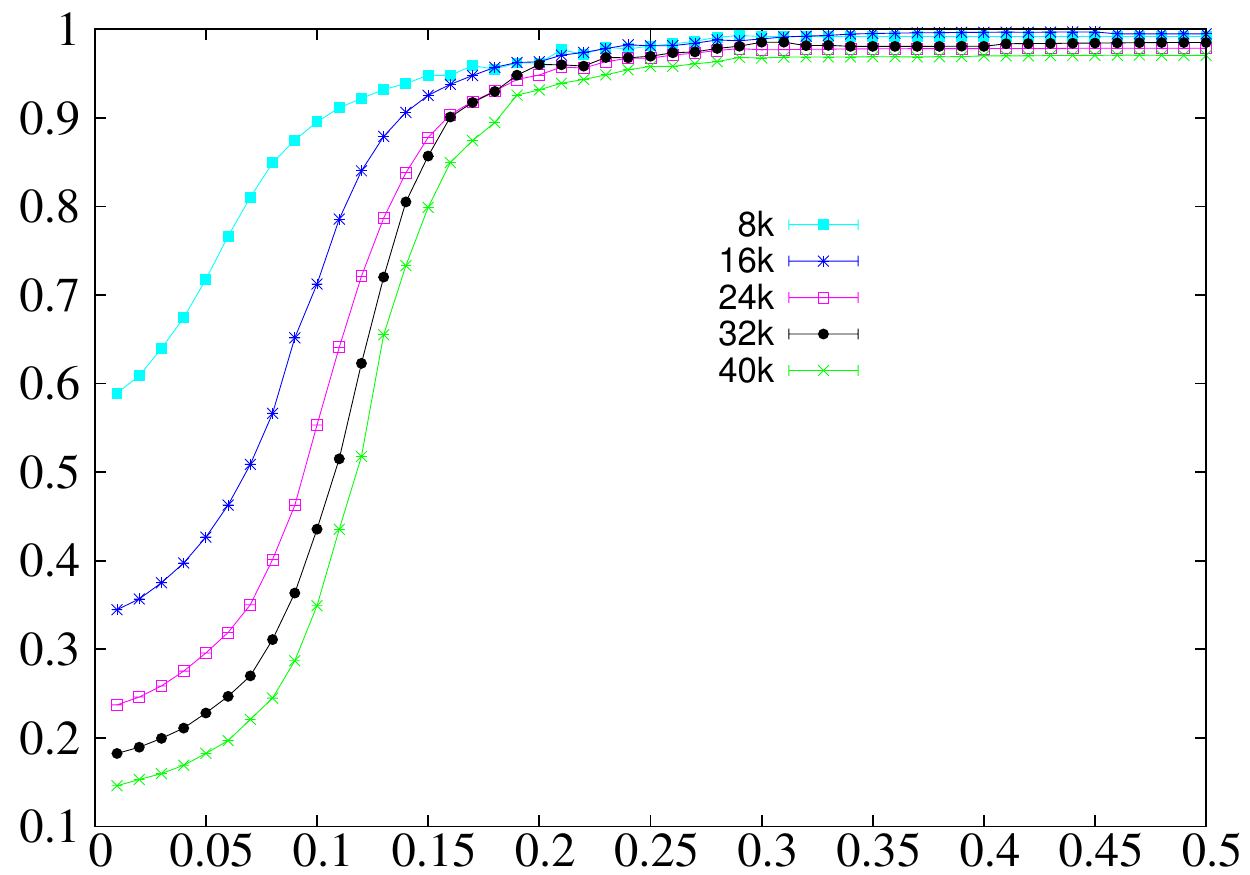}}}
\end{center}
\caption{The dependence of $ \langle \mathrm{corr}(L/2)\rangle /N$ 
and $ L \langle corr(L/2)\rangle /N^2$  on the mass in the range 
between $m^2=0.01$ and $m^2=0.30$
Both plots are for $L= 800$ with spacetime volumes $N =$ 
8000, 16000, 24000, 32000 and 40000.}
\label{hight-correlatorj}
\end{figure}

A correlator similar to that defined by (\ref{corrr}) was used in numerical
studies of the scaling in three and four-dimensional CDT \cite{correlator}.  
We will measure $\mathrm{corr}(\Delta)$ at the maximal separation $\Delta=L/2$.
In the small mass regime, where the blob is well localized we 
expect a behavior
\begin{equation}
\langle \mathrm{corr}(L/2)\rangle \approx 2 h N
\end{equation} 
where $h$ is the average spatial volume of the time slices belonging to 
the stalk. As a consequence we expect in this mass regime that 
$ \langle \mathrm{corr}(L/2)\rangle/N \approx 2h$, i.e.\  approximately 
both $N$ and $L$ independent. In the large mass regime we expect a 
different behavior
\begin{equation}
\langle \mathrm{corr}(L/2) \rangle \approx N^2/L
\end{equation}
and  consequently  $ L \langle \mathrm{corr}(L/2)\rangle /N^2\approx 1$ 
should be $N$ and $L$ independent.
In our analysis we fix the time period $L$ and the spacetime volume $N$  
and measure the correlator as a function of $m^2$. In 
Fig. \ref{hight-correlatorj}
we show the typical behavior of $ \langle \mathrm{corr}(L/2)\rangle /N$ 
and $ L \langle \mathrm{corr}(L/2)\rangle /N^2$ for $L=800$ and a
sequence of spacetime volumes $N$.  The plots illustrate the difference
between the small and large mass behavior and
indicate that there is a well defined  transition between the two regimes.

\begin{figure}[h]
\begin{center}
{\scalebox{0.48}{\includegraphics{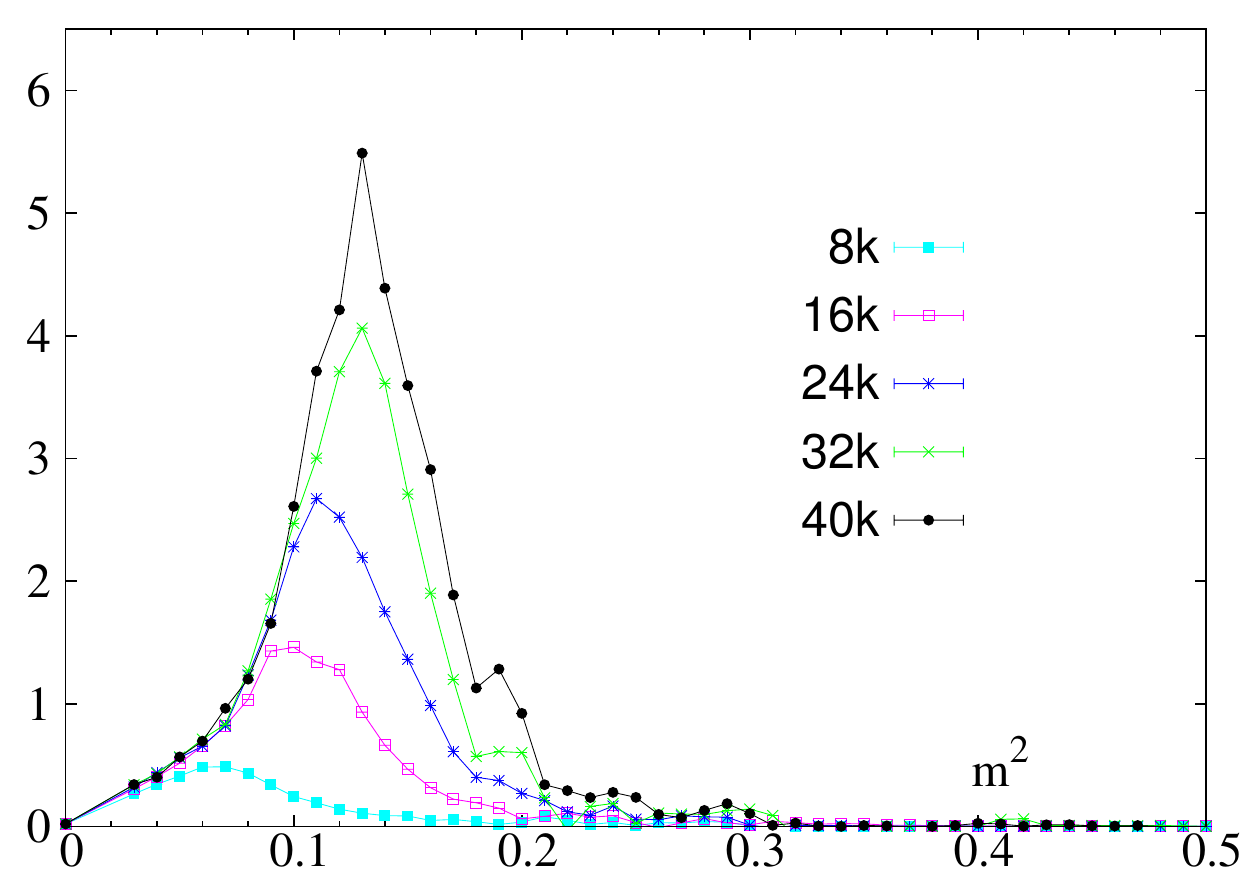}}}
\vspace{1cm}
{\scalebox{0.5}{\includegraphics{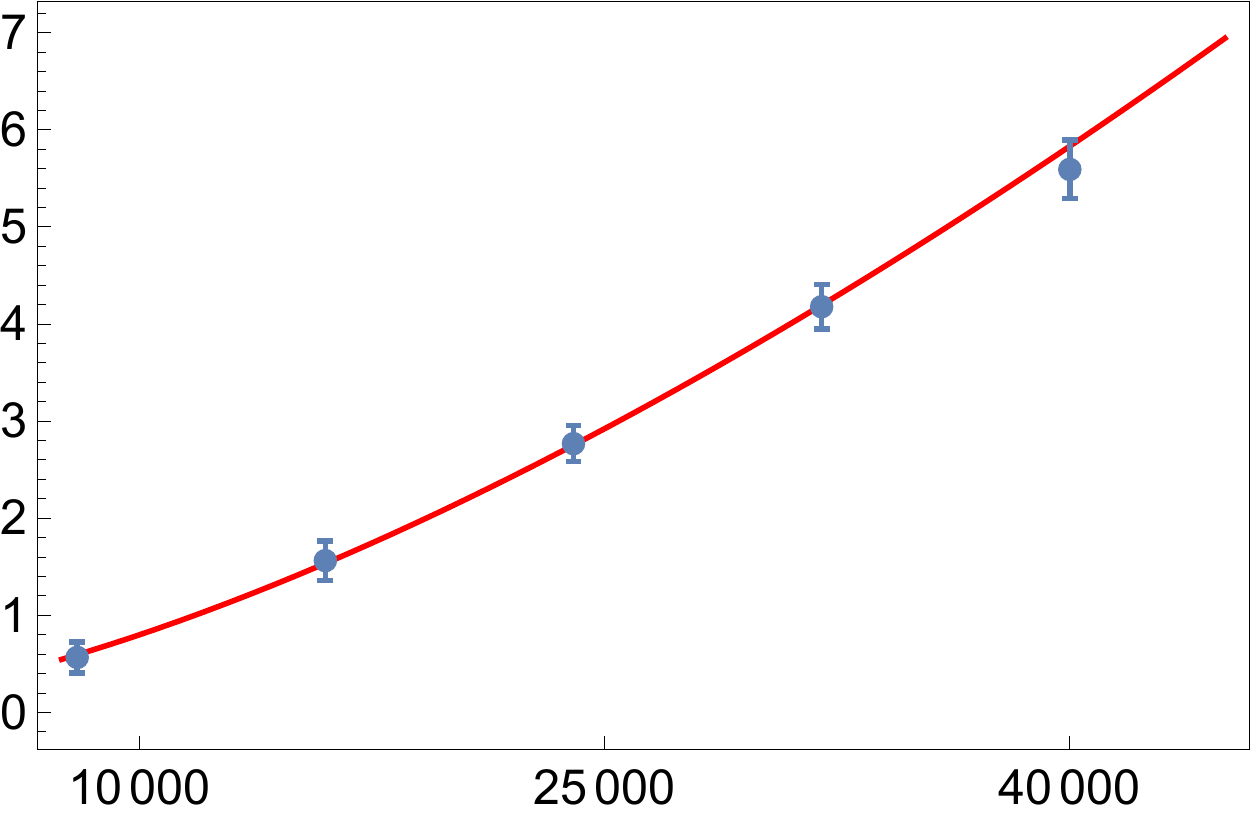}}}
\end{center}
\caption{ $(1/N) \Delta \langle \mathrm{corr}(L/2,m^2)\rangle /\Delta m^2$   
for $m^2 \in [0.01,0.30]$ (left). The curves correspond to  
$N=$ 8000, 16000, 24000, 32000 and 40000 and $L=800$. 
The right figure shows the maximum as a function of $N$, 
the curve being a fit to $N^{1.5}$.}
\label{ratio-correlator}
\end{figure}

To substantiate this we calculate 
the derivative $\d \langle \mathrm{corr}(L/2)\rangle /\d m^2$. 
It has a clear peak growing with the size $N$ of the system and thus  
signals a phase transition transition. In Fig.\ \ref{ratio-correlator} 
we show the values of the numerically estimated derivative  
$(1/N) \Delta \langle \mathrm{corr}(L/2)\rangle /\Delta m^2$ as a 
function of $m^2$ for $L=800$ and for a sequence of  spacetime volumes 
$N$ (left plot) and the peak values of the estimated derivatives as 
a function of $N$ (right plot).

The position of the maxima permits us to estimate 
the transition to be located at the critical mass 
$m^2_c \approx 0.135 \pm 0.005$. 
A more precise determination of the critical mass $m_c$ is difficult with the 
present numerical setup. 
The scaling of the maxima $H(N)$ as a function of the spacetime volume  $N$ 
can be parametrised by
\begin{equation}
H(N) \sim N^{\alpha},\quad \alpha = 1.48 \pm 0.12
\end{equation}
strongly suggesting a higher order phase transition 
(the fit is presented on right plot in Fig.\ \ref{ratio-correlator} 
as a red line). 

\section{Discussion and conclusion}\label{Conclusions}

We analyzed spatial volume distributions $\la n(t,m^2)\ra$ for CDT geometries 
interacting with 4 massive scalar fields. There seem to 
be two regimes: a small mass regime with a  universal distribution identical
to the distribution obtained for massless fields, i.e. for a conformal
field theory with central charge $c=4$, 
containing a blob and a stalk, and with the blob
scaling with  Hausdorff dimension $D_H=3$.  The other regime 
where the masses are large also has a universal distribution scaling with
$D_H=2$ and the universal distribution is the one of  
pure gravity without any matter fields. 
Using the volume-volume correlator we located the critical mass $m_c^2$ where
the transition between the two regime of different geometries takes place.
The scaling of the derivative of the correlator at the critical 
mass $m_c^2$ as a function of system size suggests that the 
phase transition is of second or higher order.

We observe the same blob structure for any number $d>1$ of massless 
Gaussian fields, as well as for multiple critical Ising spins corresponding 
to $c >1$  coupled to CDT geometries. 
We have not  observed the blobs for a single Ising 
spin, a single three-states Pott model or a single massless Gaussian field
coupled to CDT geometries, systems which all have $c \leq 1$. 
Thus it is natural to conjecture that there is $c=1$ barrier also 
in 2d CDT/Ho\v{r}ava-Lifshitz quantum gravity coupled to conformal field
theories, and that it is a transition associated with this barrier that 
we observe by changing the mass of the four Gaussian fields. 

It would be very interesting if one could solve the CDT model coupled
to Gaussian fields analytically. Understanding the $c=1$ barrier might
help us to a better understanding of the $c=1$ barrier in quantum Liouville 
gravity and understanding the formation of the blobs might help us 
to understand better the similar phenomenon in higher dimensional CDT
\cite{highd}, where the appearance of the 
blob has been important in the attempts to define
a continuum limit of lattice gravity \cite{blob,planck}.     

\vspace{1cm}

\noindent {\bf Acknowledgments.} 
HZ is partly supported by the International PhD Projects 
Programme of the Foundation for Polish Science within the 
European Regional Development Fund of the European Union, 
agreement no. MPD/2009/6. JJ acknowledges the support of grant 
DEC-2012/06/A/ST2/00389 from the National Science Centre Poland. 
JA and AG acknowledge support from the ERC-Advance grant 291092,
``Exploring the Quantum Universe'' (EQU).


\vspace{1cm}

\begin{thebibliography}{99}


\bibitem{kpz}
  V.G.~Knizhnik, A.M.~Polyakov and A.B.~Zamolodchikov,
  Mod.\ Phys.\ Lett.\ A {\bf 3} (1988) 819.
\bibitem{david}
  F.~David,
  Mod.\ Phys.\ Lett.\ A {\bf 3} (1988) 1651.
\bibitem{dk}
  J.~Distler and H.~Kawai,
  Nucl.\ Phys.\ B {\bf 321} (1989) 509.
\bibitem{david1}
  F.~David,
  Nucl.\ Phys.\ B {\bf 257 } (1985)  45.\\
A.~Billoire and F.~David,
Phys.\ Lett.\  B\ {\bf 168} (1986) 279-283.
\bibitem{kkm}
  V.A.~Kazakov, A.A.~Migdal and I.K.~Kostov,
  Phys.\ Lett.\ B {\bf 157} (1985)  295-300.\\
D.V.~Boulatov, V.A.~Kazakov, I.K.~Kostov and A.A.~Migdal,
Nucl.\ Phys.\  B\ {\bf 275} (1986) 641-686.
\bibitem{adf}
J.~Ambjorn, B.~Durhuus and J.~Fr\"ohlich,
Nucl.\ Phys.\  B\ {\bf 257} (1985) 433-449.\\
J.~Ambjorn, B.~Durhuus, J.~Fr\"ohlich and P.~Orland,
Nucl.\ Phys.\  B\ {\bf 270} (1986) 457-482.
\bibitem{horava}
  P.~Ho\v rava,
  Phys.\ Rev.\ D {\bf 79 } (2009)  084008
  [arXiv:0901.3775, hep-th].
\bibitem{agsw}
  J.~Ambjorn, L.~Glaser, Y.~Sato and Y.~Watabiki,
  Phys.\ Lett.\ B {\bf 722} (2013) 172
  [arXiv:1302.6359 [hep-th]].
\bibitem{largec}
  J.~Ambjorn, B.~Durhuus, T.~Jonsson and G.~Thorleifsson,
  Nucl.\ Phys.\ B {\bf 398} (1993) 568
  [hep-th/9208030].\\
  J.~Ambjorn and G.~Thorleifsson,
  Phys.\ Lett.\ B {\bf 323} (1994) 7
  [hep-th/9312157].\\
  J.~Ambjorn, G.~Thorleifsson and M.~Wexler,
  Nucl.\ Phys.\ B {\bf 439} (1995) 187
  [hep-lat/9411034].\\
  M.~G.~Harris and J.~Ambjorn,
  Nucl.\ Phys.\ B {\bf 474} (1996) 575
  [hep-th/9602028].\\
  F.~David,
  Nucl.\ Phys.\ B {\bf 487} (1997) 633
  [hep-th/9610037].
\bibitem{aal}
  J.~Ambjorn, K.~N.~Anagnostopoulos and R.~Loll,
  Phys.\ Rev.\ D {\bf 61} (2000) 044010
  [hep-lat/9909129].
\bibitem{agjz}
J.~Ambjorn, A.~G\"orlich, J.~Jurkiewicz and H.~G. Zhang,
Nucl.Phys. B863 (2012) 421-434, [hep-th/1201.1590].
\bibitem{al}
J.~Ambjorn and R.~Loll,
Nucl.Phys.B {\bf 536} (1998) 407-434, [hep-th/9805108].
\bibitem{smallc}
J.~Ambjorn, K.N.~Anagnostopoulos and R.~Loll:
Phys.\ Rev.\  D\ {\bf 60} (1999) 104035 [hep-th/9904012].\\
J.~Ambjorn, K.N.~Anagnostopoulos, R.~Loll and I.~Pushkina,
Nucl.\ Phys.\  B {\bf 807} (2009) 251 [arXiv:0806.3506, hep-lat].
\bibitem{planck}
J.~Ambjorn, A.~G\"orlich, J.~Jurkiewicz and R.~Loll,
Phys.\ Rev.\ Lett.\ {\bf 100} (2008) 091304 [arXiv:0712.2485, hep-th].\\
Phys.\ Rev.\  D {\bf 78} (2008) 063544 [arXiv:0807.4481, hep-th].\\
Phys.\ Lett.\ B {\bf 690} (2010) 420-426 [arXiv:1001.4581, hep-th].
\bibitem{correlator}
J.~Ambjorn, J.~Jurkiewicz and R.~Loll,
Phys.\ Rev.\ Lett.\ {\bf 93} (2004) 131301 [hep-th/0404156].\\
Phys.\ Rev.\ D\ {\bf 72} (2005) 064014  [hep-th/0505154].
Phys.\ Lett.\ B {\bf 607} (2005) 205-213
[hep-th/0411152].
\bibitem{highd}
J.~Ambjorn, J.~Jurkiewicz and R.~Loll:
Phys.\ Rev.\  D {\bf 64} (2001) 044011 [hep-th/0011276];\\
Nucl.Phys.B {\bf 610} (2001), 347-382 [hep-th/0105267].\\
Phys.\ Rev.\ Lett.\  {\bf 85} (2000) 924 [hep-th/0002050].
\bibitem{blob}
  J.~Ambjorn, A.~Goerlich, J.~Jurkiewicz and R.~Loll,
  Phys.\ Rept.\  {\bf 519} (2012) 127
  [arXiv:1203.3591 [hep-th]].\\
  J.~Ambjorn, A.~Gorlich, S.~Jordan, J.~Jurkiewicz and R.~Loll,
  Phys.\ Lett.\ B {\bf 690} (2010) 413
  [arXiv:1002.3298 [hep-th]].\\
 J.~Ambjorn, S.~Jordan, J.~Jurkiewicz and R.~Loll,
Phys.\ Rev.\ Lett.\  {\bf 107} (2011) 211303
[arXiv:1108.3932 [hep-th]].
  Phys.\ Rev.\ D {\bf 85} (2012) 124044
  [arXiv:1205.1229 [hep-th]].\\








\end{thebibliography}
\end{document}